\documentclass[12pt]{article}

\usepackage{jheppub}          
\usepackage{amsmath,amssymb}
\usepackage{graphicx}
\usepackage{bm}
\usepackage{hyperref}
\usepackage{xcolor}
\usepackage{float}
\usepackage{booktabs}
\usepackage{multirow}
\usepackage{braket}
\usepackage{subcaption}

\newcommand{\ess}{\varepsilon_{ss}}
\newcommand{\ems}{\varepsilon_{\mu s}}
\newcommand{\ms}{m_s}
\newcommand{\meff}{m^2_{\rm eff}}

\newcommand{\Vnc}{V_{\rm NC}}

\newcommand{\Psmu}{P_{s\mu}}
\newcommand{\FQ}{F_Q}
\newcommand{\FC}{F_C}
\newcommand{\QCRB}{\Delta_{\rm QCRB}}
\newcommand{\CCRB}{\Delta_{\rm CCRB}}
\newcommand{\KM}{\mathrm{KM3NeT}}
\newcommand{\IC}{\mathrm{IceCube}}
\newcommand{\eVu}{\,\mathrm{eV}}
\newcommand{\keVu}{\,\mathrm{keV}}
\newcommand{\PeVu}{\,\mathrm{PeV}}
\newcommand{\kmu}{\,\mathrm{km}}

\newcommand{\SLD}{\hat{L}}

\title{Quantum Fisher Information as a Probe of Sterile Neutrino New Physics:Geometric Advantage of KM3NeT over IceCube}

\author[]{Baktiar Wasir Farooq,}
\author[]{Bipin Singh Koranga,}
\author[]{Aritro Chatterjee}

\affiliation[]{Department of Physics, Kirori Mal College,
University of Delhi, New Delhi 110007, India}
\emailAdd{baktiarfarooq238@gmail.com}
\emailAdd{bskoranga@kmc.du.ac.in}
\emailAdd{aritro0112@gmail.com}

\abstract{
We apply Quantum Fisher Information (QFI) and Classical Fisher Information (CFI) to
quantify the precision limits on sterile neutrino new-physics parameters in the MSW and
NSI scenarios proposed to explain the KM3-230213A event. For a pure two-level neutrino
state under binary flavor projection, we prove that CFI exactly saturates the QFI for
some cases and standard muon-track detection achieves quantum-optimal sensitivity. We
further estimate the classical and quantum Cram\'er--Rao bounds for both the experiments,
demonstrating that their equality implies standard $\chi^2$ analyses performed by neutrino
telescope collaborations automatically achieve the quantum-optimal estimation bound, and
use the bounds to quantify the minimum number of events required at each detector for
equivalent parameter precision. Extending the QFI analysis to longer baselines, the MSW
scenario predicts sharp oscillatory dips at nodes where conversion probability saturates,
providing an optimal baseline of $\approx 120\,\mathrm{km}$ for IceCube and
$\approx 2500\,\mathrm{km}$ for KM3NeT for accurate estimation of both $\varepsilon_{ss}$
and $m_s$, while the NSI scenario exhibits an $L^2$-growing envelope with an optimal
baseline at $\approx 1300\,\mathrm{km}$ for estimation of $\varepsilon_{\mu s}$ for both
experiments and $\approx 2000\,\mathrm{km}$ for estimation of $m_s$, far beyond either
current detector. These results establish that the KM3NeT--IceCube tension reflects a
fundamental asymmetry in quantum information content rather than a statistical artifact,
and motivate future neutrino telescopes positioned within the quantum-sensitive baseline
window.}


\begin{document}
\maketitle

\section{Introduction}
\label{sec:intro}
The recent detection of an ultra-high-energy (UHE) neutrino event at 220~PeV by the KM3NeT Collaboration \cite{KM3NeT:2025nature} has reopened long-standing questions about physics beyond the Standard Model in the neutrino sector. This event sits in tension with two decades of IceCube observations of the astrophysical neutrino flux \cite{IceCube:2013,IceCube:2014}, and a detailed comparison of the two datasets suggests that the discrepancy cannot easily be accommodated within the standard three-flavor oscillation framework \cite{Li:2026}. Reconciling the KM3-230213A event with IceCube's non-observation of a comparable signal has therefore motivated a range of new-physics explanations, among which sterile neutrino scenarios --- mediated either by a resonant MSW-like conversion or by non-standard interactions (NSI) --- have emerged as leading candidates \cite{Brdar:2026}.

Building on the theoretical framework of Brdar and Chattopadhyay \cite{Brdar:2026}, we adopt two benchmark new-physics scenarios: an MSW resonance sourced by a new baryonic gauge interaction coupling sterile neutrinos to nucleons, parameterized by $\varepsilon_{ss}$, and an NSI scenario in which an off-diagonal matter potential $\varepsilon_{\mu s}$ directly mixes $\nu_\mu$ and $\nu_s$. The baryonic coupling underlying the MSW scenario is constrained by a body of prior work on new light gauge bosons and their associated neutral-current phenomenology \cite{Pospelov:2011,Pospelov:2012,Kopp:2014,Harnik:2012,Dror:2017}, which together set the phenomenologically allowed range of $\varepsilon_{ss}$ used throughout this analysis.

To assess how much information about these new-physics parameters is actually extractable by KM3NeT- and IceCube-like detectors, we turn to the tools of quantum estimation theory. The Quantum Fisher Information (QFI) and its associated Quantum Cram\'er--Rao Bound provide a measurement-independent benchmark for the best possible precision achievable on a parameter encoded in a quantum state \cite{Braunstein:1994,Paris:2009}, and have since been generalized to multiparameter settings \cite{Liu:2020}. This framework has recently been extended to collider physics by Ai et al.\ \cite{Ai:2025}, who asked under what conditions collider experiments --- restricted to classical observables such as reconstructed particle momenta --- can nonetheless saturate the QFI of an entangled two-particle spin state. Using the $\tau^+\tau^-$ system with the $\tau \to \pi\nu$ decay as a projective-measurement case study, they show that QFI saturation holds if and only if the symmetric logarithmic derivative commutes with a complete set of orthonormal, separable projectors accessible at colliders, a condition that in turn requires the spin density matrix to be rank-deficient. Applying this criterion, they find that the classical Fisher information asymptotically saturates the QFI for magnetic-dipole-moment and CP-violating Higgs couplings in restricted regions of phase space, but not for electric dipole moments. This result is conceptually close to the central question we ask of neutrino telescopes: whether the standard, comparatively simple measurement already performed by an experiment (flavor projection in our case, momentum reconstruction in theirs) is sufficient to reach the fundamental quantum limit, or whether some fraction of the available information is being left on the table. Closer to the present context, this quantum-estimation approach has also been applied directly to neutrino oscillation phenomenology: early work by Nogueira et al.\ examined quantum estimation in standard oscillations \cite{Nogueira:2017}, while more recent studies have used QFI to probe the origin of the large uncertainty on the CP-violating phase $\delta_{CP}$ \cite{Ignoti:2025} and to characterize parameter sensitivity in long-baseline experiments \cite{Yadav:2026}. Complementary studies have examined how non-standard interactions reshape quantum coherence and spatiotemporal correlations in propagating neutrino states \cite{Dixit:2021,Sarkar:2021}, providing additional theoretical context for treating the sterile-neutrino problem through a quantum-information lens.

This paper is also a continuation of our own prior work applying quantum-information measures to neutrino oscillation physics, including entanglement in two-flavor oscillations in matter \cite{Koranga:2025a}, entanglement-based signatures of CPT violation \cite{Koranga:2025b}, and modified entanglement structures arising from quantum-gravity-motivated interactions \cite{Koranga:2026}. The numerical machinery used here to compute oscillation amplitudes and probabilities builds on our previously developed three-flavor oscillation code \cite{Farooq:2024}, with the sterile-neutrino-specific implementation made publicly available \cite{github}.

In what follows, we compute the QFI and CFI for both the MSW and NSI sterile-neutrino scenarios as functions of baseline length for KM3NeT and IceCube, identify the conditions under which the classical flavor-projection measurement performed by neutrino telescopes saturates the quantum bound --- in the spirit of the saturation criterion developed by Ai et al.\ for collider observables \cite{Ai:2025} --- and use the corresponding Cram\'er--Rao bounds to quantify the event statistics each experiment would require to reach a given precision. We further extend this analysis to baselines well beyond the current detector geometries in order to identify optimal baseline windows for future or upgraded neutrino telescopes.

\section{Sterile neutrino oscillation framework}

\label{sec:theory}

We follow the notation and theoretical setup of
\cite{Brdar:2026} throughout this section.
For the case of the $(\nu_s, \nu_{\mu})$, the neutrinos states in qubits can be represented by:
\begin{equation}
    \ket{\nu(t)} = \mathcal{A}_{ss}(t)\ket{1}_{s}\otimes\ket{0}_{\mu}+ \mathcal{A}_{s \mu}(t)\ket{0}_{s}\otimes\ket{1}_{\mu}
    \label{eq: 2.1}
\end{equation}

Here, $\mathcal{A}_{ss}(L)$ and $\mathcal{A}_{s \mu}(L)$ denote the complex amplitudes such that $P_{ss} = |\mathcal A_{ss}(L)|^2$ is the survival probability of the sterile neutrino. In our study, we compute the complex amplitude and probability using the modified Hamiltonian for both the NSI and MSW cases computationally as in \cite{Farooq:2024}.\\

The density matrix for this state is:
\begin{equation}
\rho(t) = \begin{pmatrix}
0 & 0 & 0 & 0 \\[8pt]
0 & |\mathcal{A}_{s\mu}(L)|^2 & 
\mathcal{A}_{s\mu}(L)\,\mathcal{A}_{ss}^*(L) & 0 \\[8pt]
0 & \mathcal{A}_{ss}(L)\,\mathcal{A}_{s\mu}^*(L) & 
|\mathcal{A}_{ss}(L)|^2 & 0 \\[8pt]
0 & 0 & 0 & 0
\end{pmatrix}
\end{equation}

From the above formalism it's clear that the neutrino state propagating from source to detector is a pure state in the $(\nu_\mu, \nu_s)$ basis, evolving unitarily under the Hamiltonian of Eq.~(\ref{eq:H_MSW}) or Eq.~(\ref{eq:H_NSI})
\subsection{MSW resonance scenario}
\label{sec:MSW}

We work in the two-flavour $(\nu_\mu,\nu_s)$ basis.
The Hamiltonian governing propagation through matter of
density $\rho$ is
\begin{equation}
  H = U
  \begin{pmatrix} 0 & 0 \\ 0 & m_s^2/(2E_\nu) \end{pmatrix}
  U^\dagger
  + \begin{pmatrix} V_{\rm NC} & 0 \\ 0 & V_s \end{pmatrix} ,
  \label{eq:H_MSW}
\end{equation}
where $U$ is the $(\nu_\mu,\nu_s)$ mixing matrix with vacuum
mixing angle $\theta$, and $\Vnc=-G_F n_n/\sqrt{2}$ is the
Standard Model neutral-current potential.  The sterile neutrino
acquires a new baryonic matter potential
\begin{equation}
  V_s = \sqrt{2}\,G_F\,\ess\,(n_p+n_n)
        \approx 2\,\ess\,V_{\rm CC},
  \label{eq:Vs}
\end{equation}
where $\ess = G_B/(\sqrt{2}\,G_F)$ parameterizes the new baryonic
interaction strength relative to the Fermi constant.  In the
literature, $|\ess|$ values up to $\mathcal{O}(10^2$--$10^3)$
have been explored~\cite{Pospelov:2011,Pospelov:2012,Kopp:2014},
with the maximum phenomenologically allowed value
$|\ess|\lesssim450$ set by anomaly-cancellation constraints
on the associated $U(1)_B$ vector boson~\cite{Harnik:2012,Dror:2017}.

MSW resonance occurs when
$V_s = (m_s^2/2E_\nu)\cos2\theta$.
For the 220\,PeV event and matter density
$\bar\rho\approx1.82\,\mathrm{g\,cm}^{-3}$ (averaged over the
KM3NeT Earth path, comprising 100\,km of rock at
$2.2\,\mathrm{g\,cm}^{-3}$ and 47\,km of sea water at
$1\,\mathrm{g\,cm}^{-3}$), resonance occurs at
$\sqrt{\Delta m^2}\approx m_s\simeq2\times10^{-1}\sqrt{|\ess|}\keVu$.
The benchmark \emph{resonance parameter} identified in literature (Supplementary Material \cite{Brdar:2026})is given by 
\begin{equation}
  \ess = -150, \quad m_s = 3\keVu,
  \label{eq:benchmark_MSW}
\end{equation}
at which the MSW peak in $\Psmu(E_\nu)$ falls in the 150--300\,PeV
window and $\Psmu$ for KM3NeT is two orders of magnitude larger
than for IceCube~\cite{Brdar:2026}.  We adopt these benchmark
values throughout; for completeness we note that the resonance
condition $V_s = (m_s^2/2E_\nu)\cos2\theta$ gives
$\ess \approx -m_s^2/(4E_\nu V_{\rm CC}) \approx -150$ for
$m_s=3\keVu$, $E_\nu=220\PeVu$.

The effective matter-modified oscillation parameters are
\begin{align}
  \meff &= \sqrt{m_s^4 + C^2\ess^2 + d\,\ess},
  \label{eq:meff} \\
  \sin^2(2\theta_m) &= \frac{m_s^4\sin^2 2\theta}{(\meff)^2},
  \label{eq:sin2}
\end{align}
with the shorthand $C=4E_\nu V_{\rm CC}$ and
$d=8m_s^2 E_\nu V_{\rm CC}\cos2\theta$.
The sterile-to-muon conversion probability after baseline $L$ is
\begin{equation}
  \Psmu^{\rm MSW}(L) = \sin^2(2\theta_m)\,
    \sin^2\!\left(\frac{1.27\,\meff\,L}{E_\nu}\right),
  \label{eq:Psmu_MSW}
\end{equation}
with $E_\nu$ in GeV, $\meff$ in $\mathrm{eV}^2$, and $L$ in km.

\subsection{Non-standard interaction scenario}
\label{sec:NSI}

In the NSI scenario an off-diagonal matter potential couples
$\nu_\mu$ and $\nu_s$, modifying Eq.~\eqref{eq:H_MSW} to
\begin{equation}
  H = U
  \begin{pmatrix} 0 & 0 \\ 0 & m_s^2/(2E_\nu) \end{pmatrix}
  U^\dagger
  + \begin{pmatrix} V_{\rm NC} & \ems V_{\rm CC} \\
                    \ems V_{\rm CC} & 0 \end{pmatrix}.
  \label{eq:H_NSI}
\end{equation}
\\
This off-diagonal term, arising from the effective Lagrangian 
\begin{center}
    \vspace{0.5cm}
$\mathcal{L}\supset-2\sqrt{2}\,G_F\,\ems^f
(\bar\nu_s\gamma^\mu P_L\nu_\mu)(\bar{f}\gamma_\mu f)$, 
\vspace{0.5cm}
\end{center}
 does not involve MSW resonance but produces an
$L^2$-enhanced oscillation probability in the regime
$m_{\rm eff}^2 L/(4E_\nu)\lesssim\mathcal{O}(1)$.
In the limit of small vacuum mixing $\theta$ and
$E_\nu V_{\rm CC}\gg m_s^2$, the effective parameters reduce to :-
\begin{equation}
\theta_m^{\rm NSI} = \frac{1}{2}\tan^{-1}\left(\frac{4\ems E_\nu V_{cc}}{m_s^2+E_\nu V_{cc}}\right),
\label{eq:theta_NSI}
\end{equation}
\begin{equation}
m_{\rm eff}^{2,{\rm NSI}} = \sqrt{(E_\nu V_{\rm CC}+m_s^2)^2
    + 16\ems^2 E_\nu^2 V_{\rm CC}^2}
    \label{eq:m^2eff_NSI}
\end{equation}
and the conversion probability is
\begin{equation}
  \Psmu^{\rm NSI}(L) = \sin^2(2\theta_m^{\rm NSI})\,
    \sin^2\!\left(\frac{m_{\rm eff}^{2,{\rm NSI}}\,L}{4E_\nu}\right).
  \label{eq:Psmu_NSI}
\end{equation}
The benchmark parameters for the NSI scenario are
$\ems=1$, $m_s=500\eVu$, $\theta=10^{-4}$.  In the $L^2$ regime,
the ratio of conversion probabilities scales as
$(L_{\KM}/L_{\IC})^2\approx110$, independent of $\ems$.

\section{Fisher information framework}
\label{sec:QFI}

\subsection{Fisher Information: definition}
\label{sec:FI_def}
Fisher Information which is a subset of the Estimation Theory measures the  amount of information carried by an observable distribution about an unknown parameter that models it.  \cite{Nogueira:2017}

\subsection{Classical and Quantum Fisher Information for flavor projection}
\label{sec:CFI and QFI_def}

Classical Fisher Information (CFI) provides measurement oriented information about the unknown parameter ($\lambda$), it tells how much information about $\lambda$ is associated with the measurement technique. In this case, we detect the active neutrinos ($\nu_{\mu}$) via detector experiments - Km3NeT and IceCube about the unknown parameter $\lambda$. for such case, the CFI is defined by:

\begin{equation}
F_C(\lambda) = \sum_{\alpha = \mu,s} \frac{1}{P_{s \to \alpha}} 
\left(\frac{\partial P_{s \to \alpha}}{\partial \lambda}\right)^2
\end{equation}

For the binary flavor-projection measurement
$\{\Pi_\mu = |\nu_\mu\rangle\langle\nu_\mu|,\,
   \Pi_s = |\nu_s\rangle\langle\nu_s|\}$
with outcome probabilities $\Psmu$ and $1-\Psmu$, the CFI is
\begin{equation}
  \FC(\lambda) = \frac{1}{\Psmu(1-\Psmu)}
    \left(\frac{\partial\Psmu}{\partial\lambda}\right)^2.
  \label{eq:CFI_binary}
\end{equation}

For a quantum state $\rho(\lambda)$ depending on a parameter
$\lambda$, the QFI is defined as~\cite{Braunstein:1994,Paris:2009}
\begin{equation}
  \FQ[\rho,\lambda] = \mathrm{Tr}\!\left[\rho\,\SLD^2\right],
  \label{eq:QFI_def}
\end{equation}
where the symmetric logarithmic derivative (SLD) $\SLD$ satisfies
the operator equation
\begin{equation}
  \partial_\lambda\rho = \tfrac{1}{2}(\rho\,\SLD + \SLD\,\rho).
  \label{eq:SLD}
\end{equation}
For a \emph{pure state} $\rho=|\psi(\lambda)\rangle\langle\psi(\lambda)|$,
the QFI reduces to the simple form~\cite{Liu:2020}
\begin{equation}
  \FQ(\lambda) = 4\!\left[\langle\partial_\lambda\psi|
    \partial_\lambda\psi\rangle
    - |\langle\partial_\lambda\psi|\psi\rangle|^2\right].
  \label{eq:QFI_pure}
\end{equation}
The QFI sets the ultimate precision on $\lambda$, irrespective of
measurement strategy, via the Quantum Cramer Rao Bound (QCRB) :
\begin{equation}
  \Delta\lambda \geq \frac{1}{\sqrt{N\,\FQ(\lambda)}}.
  \label{eq:QCRB}
\end{equation}

By the data-processing inequality, $\FC(\lambda)\leq\FQ(\lambda)$
in general.\\
So, for a particular state at a given time it's not possible to achieve information about the parameter associated with it more than the QFI as indicated in the above equation, so we try to study the cases and their respective saturation condition i.e when QFI = CFI for a single event, then the measurement provides the most accurate estimation of the parameter, leading to a better study.

\subsection{QFI and CFI for the MSW scenario}
\label{sec:QFI_MSW}
\label{sec:eps_ss}
For $\lambda=$ the estimation parameter, the respective QFI and CFI turns out to be :
\begin{equation}
    F_Q(\lambda;L) = 4\left||\partial_{\lambda}A_{ss}|^2 + |\partial_{\lambda}A_{s\mu}|^2 -|A_{ss}(\partial_{\lambda}{A_{ss}^{*}}) + A_{s\mu}(\partial_{\lambda}{A_{s\mu}^{*}})| \right|^2 
\end{equation}

 Here, $A_{ss}\equiv A_{ss}(L)$ and $A_{s\mu}\equiv A_{s\mu}(L)$

\begin{equation}
   \FC(\epsilon_{ss};L) = \frac{(\partial_{\epsilon_{ss}}\Psmu)^2}
    {\Psmu(1-\Psmu)}.
  \label{eq:FQ_formula}
\end{equation}
The QFI and CFI is represented as $F_Q(\lambda,L)$ and $F_C(\lambda,L)$, this means that we are trying to study the efficiency of the parameter $\lambda$ w.r.t the baseline length (L) for the $\ket{\nu_{s}}$.\\
In these two equations, we compute the values of $\mathcal{A}_{ss}$ and $\mathcal{A}_{s\mu}$ as listed in the github repository \cite{github}.
\begin{equation}
    A_{\alpha \rightarrow\beta} = \bra{\nu_{\beta}}e^{(-i\hat{H}L)}\ket{\nu_{\alpha}}
\end{equation}

We further use \textit{np.linalgh.eigh} feature of the \textit{numpy} module to solve the Hamiltonian.
Figures \ref{fig:FQ_KM3_MSW_ess} - \ref{fig:FQ_IC_NSI_ms} shows
$\FQ(\lambda)$ and $\FC(\lambda)$ as functions of baseline $L$ for
KM3NeT and IceCube.
\subsubsection{For $\lambda = \epsilon_{ss}$} 
\begin{figure}[H]
  \centering
  \includegraphics[width=0.9\columnwidth]{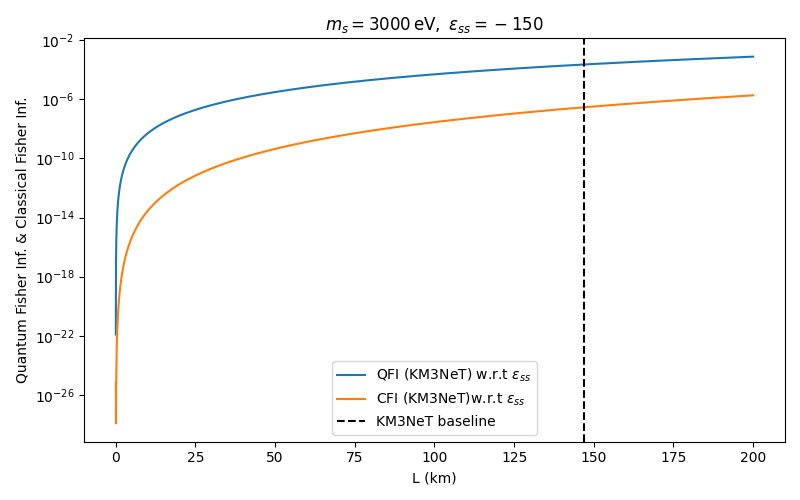}
  \caption{Quantum Fisher Information $\FQ(\ess;L)$ (blue) and
    Classical Fisher Information $\FC(\ess)$ (orange) as functions
    of baseline $L$, for \textbf{KM3NeT}, evaluated at
    $\ess=-150$, $\ms=3\keVu$, $\theta=10^{-2}$,
    $E_\nu=220\PeVu$.
    The baseline length is - $L_{\KM}=147\kmu$.}
  \label{fig:FQ_KM3_MSW_ess}
\end{figure}

From the Fig. \ref{fig:FQ_KM3_MSW_ess} it is evident that the graph strictly follows $F_Q > F_C$. To be precise, $F_Q \approx 2.6 \times 10^{-4}$ and $F_C \approx 3.5 \times 10^{-7}$ meaning that for the study of sterile neutrino  state, KM3NeT loses information about the $\nu_s$ state by 3 orders. 

\begin{figure}[H]
  \centering
  \includegraphics[width=0.9\columnwidth]{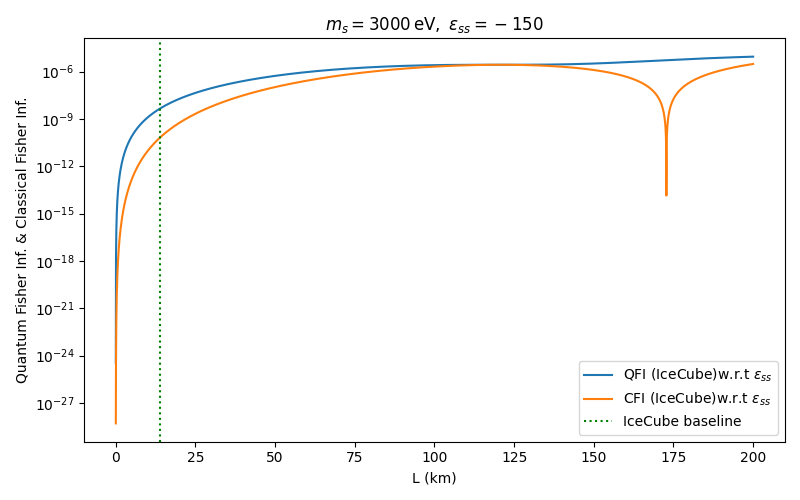}
  \caption{Quantum Fisher Information $\FQ(\ess;L)$ (blue) and
    Classical Fisher Information $\FC(\ess)$ (orange) as functions
    of baseline $L$, for \textbf{IceCube}, evaluated at
    $\ess=-150$, $\ms=3\keVu$, $\theta=10^{-2}$,
    $E_\nu=220\PeVu$.
    The baseline length is - $L_{\IC}=14\kmu$.}
  \label{fig:FQ_IC_MSW_ess}
\end{figure}
The Fig. \ref{fig:FQ_IC_MSW_ess} indicates that the graph follows $F_Q \geq F_C$. To be precise, $F_Q \approx 4.5 \times 10^{-9}$ and $F_C \approx 7.4 \times 10^{-11}$ implying that for the study of the sterile neutrino state, IceCube loses information about the $\nu_s$ state by $\approx$ 2 orders. This means that we can have better accuracy about the estimation of $m_s$ than $\epsilon_{ss}$, but the information contained within the state is less than that of $F_Q (\epsilon_{ss};L)$. Furthermore, we also observe that QFI = CFI for baseline L = 120 km, suggesting that for the given matter and sterile neutrino potential for IceCube, we will have all the information about the sterile neutrino state in case we manage to form a baseline of 120 km.

\subsubsection{For $\lambda = {m_s}$} \label{sec:m_s_MSW}

\begin{figure}[H]
  \centering
  \includegraphics[width=0.9\columnwidth]{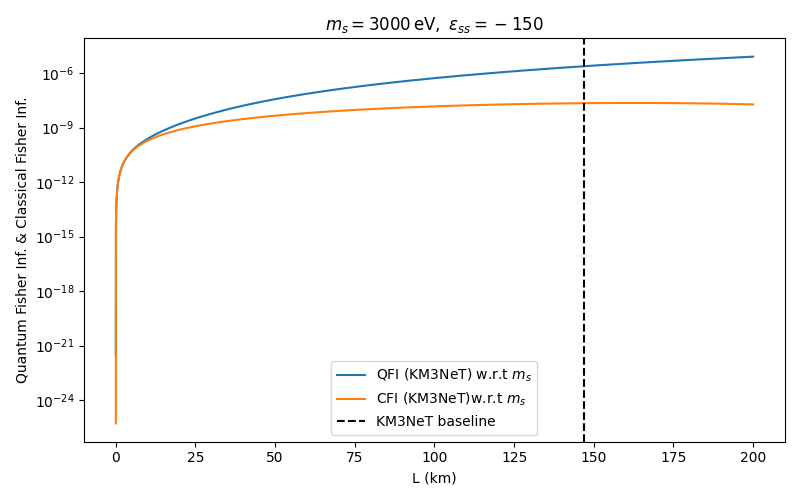}
  \caption{$\FQ(\ms)$ and $\FC(\ms)$ vs.\ $L$ for
    \textbf{KM3NeT} (MSW scenario).
    Parameters as in Fig.~\ref{fig:FQ_KM3_MSW_ess}.
    The absolute scale is smaller than for $\ess$ by
    several orders of magnitude, reflecting the reduced
    sensitivity of $\Psmu$ to small variations in $\ms$
    at fixed energy.}
    \label{fig:FQ_KM3_MSW_ms}
\end{figure}

From Fig. \ref{fig:FQ_KM3_MSW_ms} it is evident that the graph strictly follows $F_Q \geq F_C$. We can observe that the information about $\ket{\nu_{s}}$ is nearly negligible for $L<16$ km, now on increasing the baseline length, we observe that $F_C$ continues to decrease compared to $F_Q$ and with time, the L becomes approximately equal to 147 km, the $F_Q \approx 2.5 \times 10^{-6}$ and $F_C \approx 2.5 \times 10^{-8}$ meaning that for the study of the sterile neutrino state, KM3NeT loses information about the $\nu_s$ state by $\approx$ 2 orders. 

\begin{figure}[H]
  \centering
  \includegraphics[width=0.9\columnwidth]{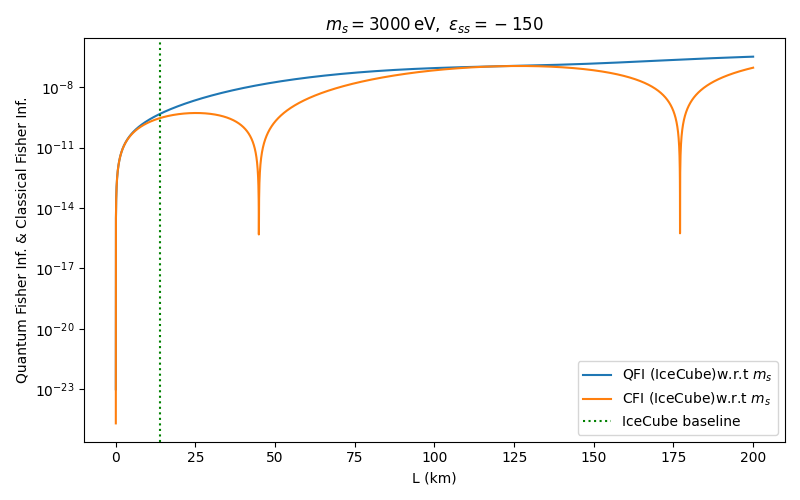}
  \caption{$\FQ(\ms)$ and $\FC(\ms)$ vs.\ $L$ for
    \textbf{IceCube} (MSW scenario).
    IceCube's 14\,km baseline lies below the sensitive region
    $L\gtrsim120\kmu$.}
  \label{fig:FQ_IC_MSW_ms}
\end{figure}
From Fig. \ref{fig:FQ_IC_MSW_ms} it is evident that the graph strictly follows $F_Q \geq F_C$. We observe an oscillation pattern for $F_C$, i.e for a $\ket{\nu_{s}}$, on traveling from 0 km to 200 km - for  $L\approx$ 50 km, 175 km, $Q_C$ drastically decreases, implying that for such baselines, experiments will not be able to capture better estimation of the parameter ($m_s$). With IceCube in its current state for L = 14 km, we observe that $F_Q \approx 4.96 \times 10^{-10}$ and $F_C \approx 3.05 \times 10^{-10}$ meaning that negligible information about the parameter is lost. We observe that for for $L = 120$ km, $F_Q = F_C$ which means that if the baseline for IceCube would have been 120 Km, it would have extracted all the information about the dependence of $\ket{\nu_s}$ on $m_s$.\\

We observe that IceCube has minimal difference between the Classical and Quantum Fisher Information in comparison to Km3NeT, which raises the question - Why could KM3NeT experimentally capture the sterile neutrino event \cite{Brdar:2026} but not IceCube? We observe that the $F_Q$ for both the parameters ($m_s,\epsilon_{ss}$) is always less for IceCube than KM3NeT. This solution is exclusive to the MSW analysis.

\subsection{QFI and CFI for the NSI scenario}
\label{sec:QFI_NSI}
For $\lambda=$ the estimation parameter, the respective QFI and CFI turns out to be :
\begin{equation}
    F_Q(\lambda;L) = 4\left||\partial_{\lambda}A_{ss}|^2 + |\partial_{\lambda}A_{s\mu}|^2 -|A_{ss}(\partial_{\lambda}{A_{ss}^{*}}) + A_{s\mu}(\partial_{\lambda}{A_{s\mu}^{*}})| \right|^2 
\end{equation}

 Here, $A_{ss}\equiv A_{ss}(L)$ and $A_{s\mu}\equiv A_{s\mu}(L)$

\begin{equation}
   \FC(\epsilon_{\ems};L) = \frac{(\partial_{\epsilon_{\ems}}\Psmu)^2}
    {\Psmu(1-\Psmu)}.
  \label{eq:FQ_formula}
\end{equation}
The QFI and CFI is represented as $F_Q(\lambda,L)$ and $F_C(\lambda,L)$, this means that we are trying to study the efficiency of the parameter $\lambda$ w.r.t the baseline length (L) for the $\ket{\nu_{s}}$.\\
In these two equations, we compute the values of $\mathcal{A}_{ss}$ and $\mathcal{A}_{s\mu}$ as listed in the github repository \cite{github}.
\begin{equation}
    A_{\alpha \rightarrow\beta} = \bra{\nu_{\beta}}e^{(-i\hat{H}L)}\ket{\nu_{\alpha}}
\end{equation}

We further use \textit{linalgh} feature of the \textit{numpy} module to solve the Hamiltonian.
Figures \ref{fig:FQ_KM3_MSW_ess} - \ref{fig:FQ_IC_NSI_ms} shows
$\FQ(\lambda)$ and $\FC(\lambda)$ as functions of baseline $L$ for
KM3NeT and IceCube.

\subsubsection{For $\lambda=\ems$}\label{sec:eps_mus}

\begin{figure}[H]
  \centering
  \includegraphics[width=0.9\columnwidth]{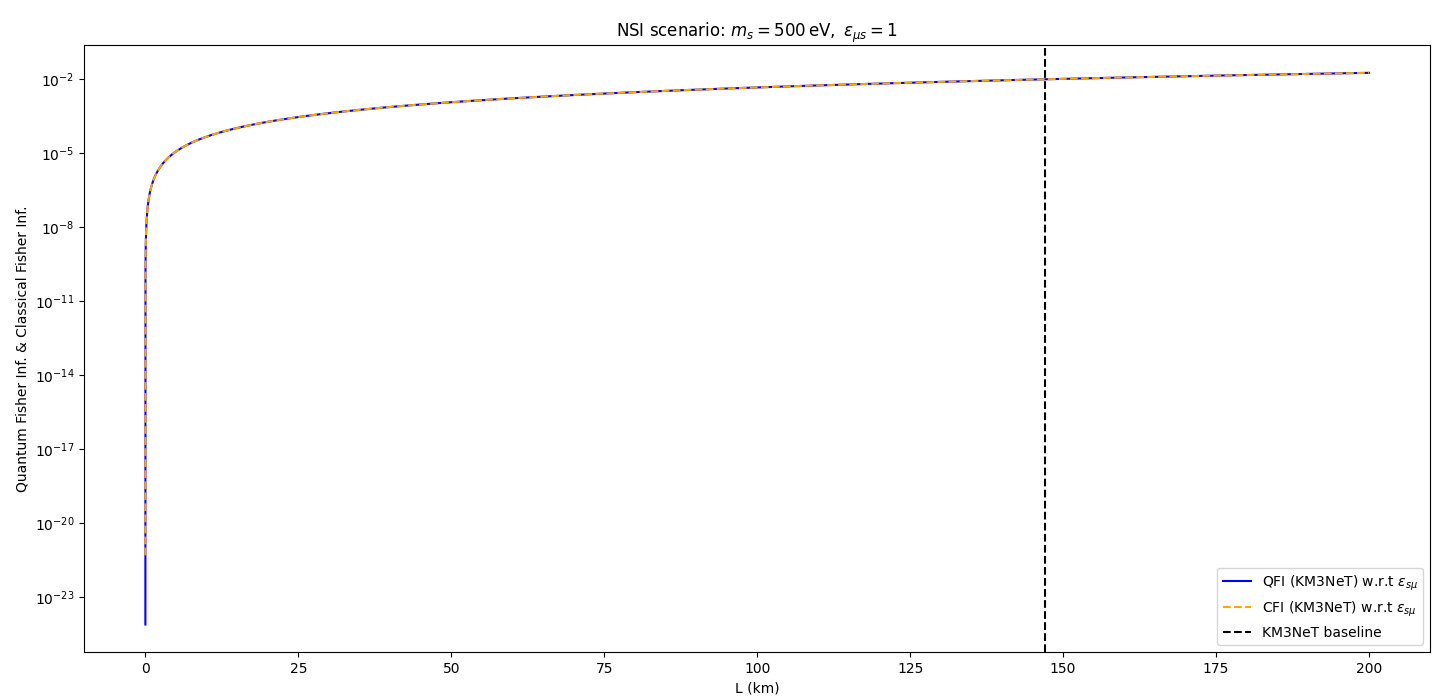}
  \caption{Quantum Fisher Information $\FQ(\ems;L)$ (blue) and
    Classical Fisher Information $\FC(\ems)$ (orange) as functions
    of baseline $L$, for \textbf{KM3NeT}, evaluated at
    $\ems=1$, $\ms=500\eVu$, $\theta=10^{-4}$,
    $E_\nu=220\PeVu$.
    The baseline length is - $L_{\KM}=147\kmu$.}
  \label{fig:FQ_KM3_NSI_ems}
\end{figure}
From Fig. \ref{fig:FQ_KM3_NSI_ems} it can be deduced that $F_Q = F_C$. To be precise, $F_Q = F_C \approx 9.46 \times 10^{-3}$, implying that for the study of sterile neutrino state $\ket{\nu_s}$, there is no information loss about the parameter $\ems$ for Km3Net.

\begin{figure}[H]
  \centering
  \includegraphics[width=0.9\columnwidth]{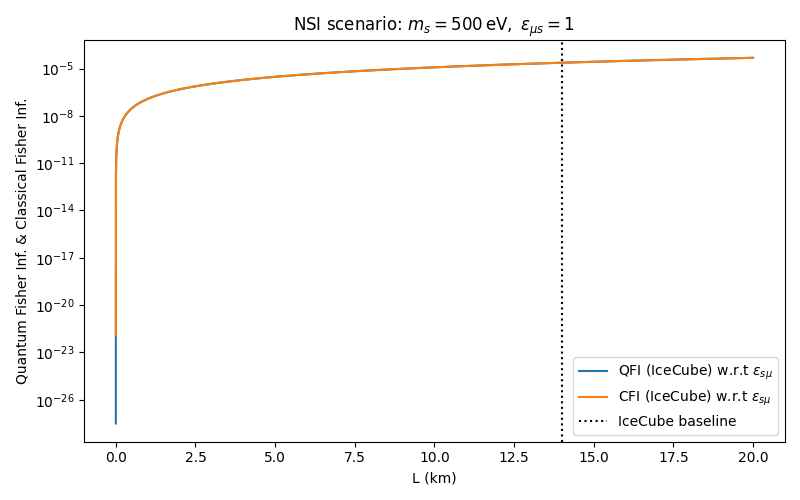}
  \caption{Quantum Fisher Information $\FQ(\ems;L)$ (blue) and
    Classical Fisher Information $\FC(\ems)$ (orange) as functions
    of baseline $L$, for \textbf{IceCube}, evaluated at
    $\ems=1$, $\ms=500\eVu$, $\theta=10^{-4}$,
    $E_\nu=220\PeVu$.
    The baseline length is - $L_{\KM}=14\kmu$.}
  \label{fig:FQ_IC_NSI_ems}
\end{figure}
Fig. \ref{fig:FQ_IC_NSI_ems} again indicates that $F_Q = F_C$. Precisely, $F_Q = F_C \approx 2.05 \times 10^{-5}$, implying that for the study of sterile neutrino state $\ket{\nu_s}$, there is no information loss about the parameter $\ems$ for IceCube.

\subsubsection{For $\lambda=m_s$}\label{sec:m_s_NSI}

\begin{figure}[H]
  \centering
  \includegraphics[width=0.8\columnwidth]{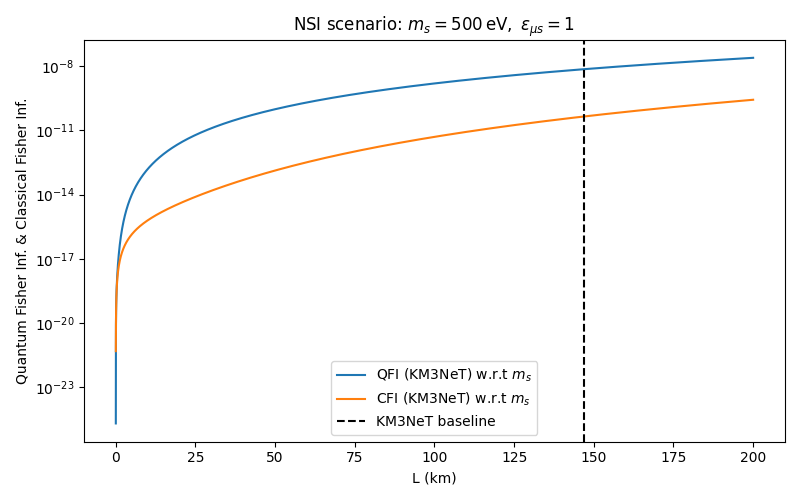}
  \caption{Quantum Fisher Information $\FQ(m_s;L)$ (blue) and
    Classical Fisher Information $\FC(m_s)$ (orange) as functions
    of baseline $L$, for \textbf{Km3Net}, evaluated at
    $\ems=1$, $\ms=500\eVu$, $\theta=10^{-4}$,
    $E_\nu=220\PeVu$.
    The baseline length is - $L_{\KM}=147\kmu$.}
  \label{fig:FQ_KM3_NSI_ms}
\end{figure}
From Fig. \ref{fig:FQ_KM3_NSI_ms} it is evident that the graph strictly follows $F_Q \geq F_C$. To be precise, $F_Q \approx 8.07 \times 10^{-9}$ and $F_C \approx 5.06 \times 10^{-11}$ meaning that for the study of sterile neutrino state, KM3NeT loses information about the $\nu_s$ state by $\approx$ 2 orders.

\begin{figure}[H]
  \centering
  \includegraphics[width=0.8\columnwidth]{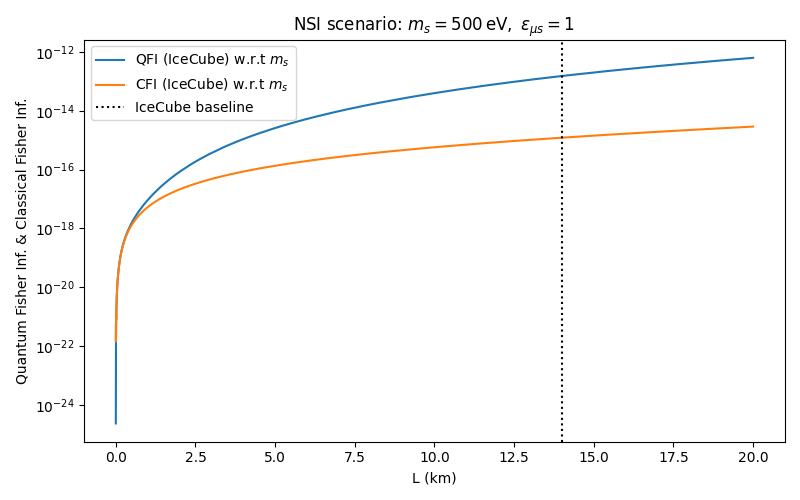}
  \caption{Quantum Fisher Information $\FQ(m_s;L)$ (blue) and
    Classical Fisher Information $\FC(m_s)$ (orange) as functions
    of baseline $L$, for \textbf{IceCube}, evaluated at
    $\ems=1$, $\ms=500\eVu$, $\theta=10^{-4}$,
    $E_\nu=220\PeVu$.
    The baseline length is - $L_{\KM}=14\kmu$.}
  \label{fig:FQ_IC_NSI_ms}
\end{figure}
From Fig. \ref{fig:FQ_IC_NSI_ms} it is evident that the graph strictly follows $F_Q \geq F_C$. To be precise, $F_Q \approx 1.73 \times 10^{-13}$ and $F_C \approx 1.21 \times 10^{-15}$ meaning that for the study of sterile neutrino state, IceCube loses information about the $\nu_s$ state by $\approx$ 2 orders. \\

We observe that for the $\lambda=\ems$ parameter, both IceCube and Km3NeT have $F_Q=F_C$ whereas for the $\lambda=m_s$ parameter, there is an information loss of 2 orders for both IceCube and Km3NeT. This implies that we have better accuracy about the estimation of $\ems$ than $m_s$ for the sterile neutrino state $\ket{\nu_s}$. We observe that $F_Q$ for both parameters ($m_s,\ems$) is always less for IceCube than KM3NeT. This solution is exclusive to the NSI scenario.

\section{Cram\'er--Rao bound fraemwork}
\label{sec:CRB}
\subsection{Cram\'er--Rao bound: definition}
The Cram\'er--Rao bound is a fundamental result in estimation theory that sets a lower limit on the variance (uncertainty) of any unbiased estimator of a parameter $\lambda$, in terms of the Fisher information $F(\lambda)$
\begin{equation}
    Var(\lambda) \geq \frac{1}{F(\lambda)}
\end{equation}

So more the Fisher information about a parameter, the tighter (smaller) is the best-possible variance on our estimate of that parameter. As long as unbiased estimators are considered, the Cram\'er Rao Bound sets the precision limit for the parameter. 
In our work we use two types of the CRB : Quantum Cram\'er Rao Bound (QCRB) and Classical Cram\'er Rao Bound (CCRB).
\subsubsection{Quantum Cram\'er--Rao bound}
\label{sec:QCRB}

The QCRB establishes the minimum achievable variance on any
unbiased estimator of $\lambda = \epsilon_{ss},m_s,\epsilon_{s\mu}$ from $X =N$ independent events:
\begin{equation}
  \Delta\lambda \geq \QCRB(\lambda;X)
    \equiv \frac{1}{\sqrt{N\,\FQ(\lambda;X)}}.
  \label{eq:QCRB_def}
\end{equation}
At $N=1$ (single-event, corresponding to KM3-230213A event captured by KM3NeT),
\begin{equation}
  \QCRB^{(1)}(\lambda;X) = \frac{1}{\sqrt{\FQ(\lambda;X)}}.
\end{equation}

\subsubsection{Classical Cram\'er--Rao bound}
\label{sec:CCRB}

For a classical estimator based on the observed flavor
(muon versus non-muon) at $N$ events, the classical
Cram\'er--Rao bound is
\begin{equation}
  \Delta\lambda \geq \CCRB(\lambda;X)
    \equiv \frac{1}{\sqrt{N\,\FC(\lambda;X)}}.
  \label{eq:CCRB_def}
\end{equation}

In Sections \ref{sec:value_of_QFI} and \ref{sec:value_of_QFI_NSI}, we use the idea of QCRB and CCRB for neutrino oscillation analysis in both the KM3NeT and IceCube experiments to understand the precision of the new physics parameters ($\lambda = \epsilon_{ss},m_s, \epsilon_{s\mu}$), leading to the capture of the oscillated sterile neutrinos to muon neutrino by varying the number of events for each of the experiments.

This analysis provides a quantitative assessment of the number of events required in each experiment to achieve a desired level of precision while simultaneously estimating the uncertainty associated with the current determination of neutrino oscillation parameters.

\subsection{CRB for MSW scenario}\label{sec:value_of_QFI}
\begin{figure}[H]
  \centering
  \includegraphics[width=0.9\columnwidth]{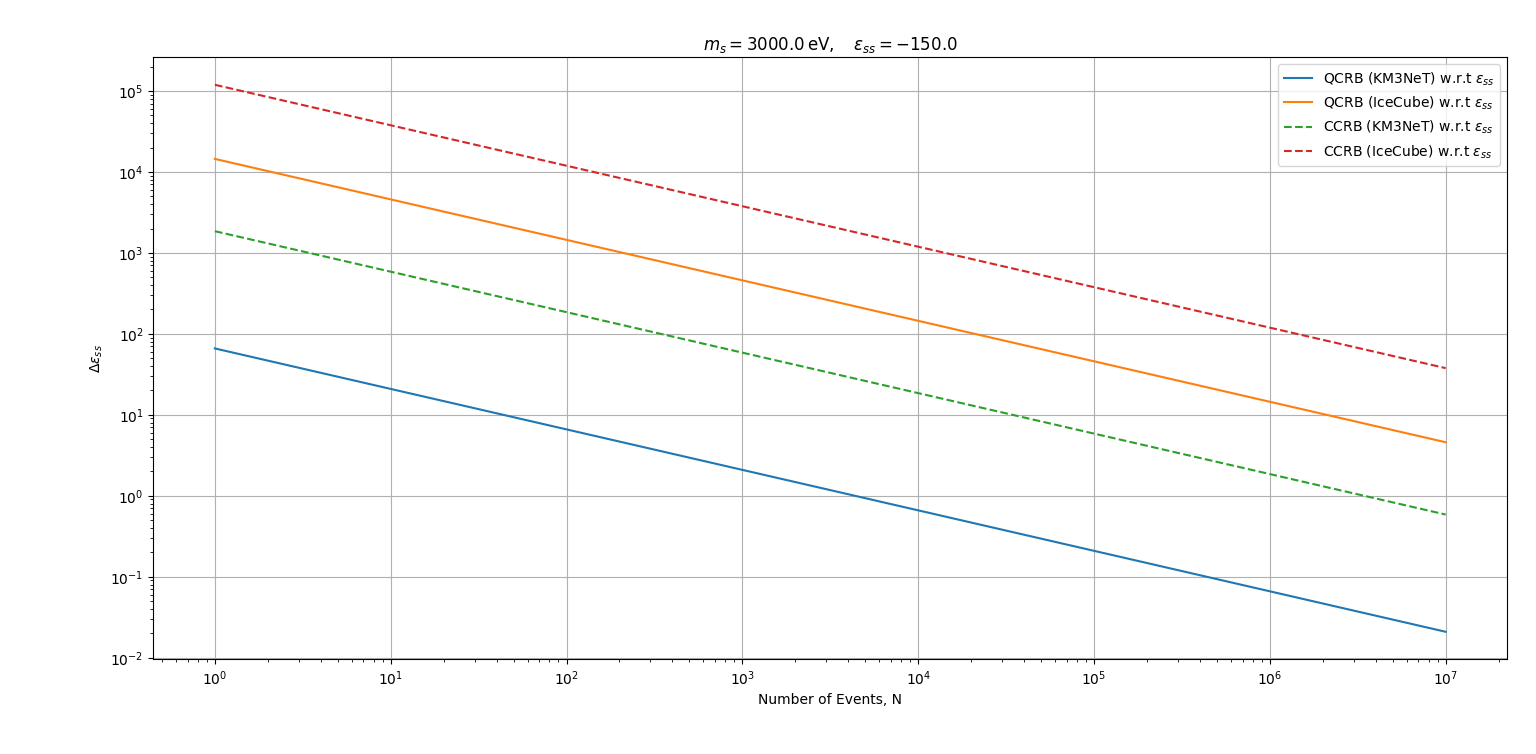}
  \caption{$\Delta \epsilon_{ss}$ vs.\ No. of events $(N)$ for
    \textbf{KM3NeT and IceCube} (MSW scenario).
    Parameters as in Fig.~\ref{fig:FQ_KM3_MSW_ess}.
    The absolute scale is smaller than for $\ess$ by
    several orders of magnitude, reflecting the reduced
    sensitivity of $\Psmu$ to small variations in $\ms$
    at fixed energy and baseline length (L).}
  \label{fig:CRB_MSW_eps_ss}
\end{figure}

It is evident from the plots that as the events capture numbers increase, the uncertainty associated with the parameters steadily decreases. Talking about Figure \ref{fig:CRB_MSW_eps_ss}, we have set the baseline length to KM3NeT ($\approx 147$ km) and IceCube ($\approx 14$ km) and carried out our study (QCRB and CCRB) by varying the Number of events (N) for both the experiments. \\

For $\lambda = \epsilon_{ss}$,\\

For KM3NeT,
\begin{itemize}
    \item For the current scenario (i.e, N = 1), the CCRB is observed to be $\approx1908.89$ and the respective QCRB is observed to be $\approx61$, this ensures that as these observations have been encoded with such large uncertainties so for more accurate analysis, the event number must be increased: On increasing the event numbers we observe that the uncertainty, the uncertainty linearly decreases. For $10^7$ events, the CCRB drops to $\approx0.62$ and the respective QCRB drops to $\approx0.02$, this shows the enormous drop in uncertainty.
\end{itemize}
For IceCube,
\begin{itemize}
    \item For the current scenario (i.e, N = 1), the CCRB is observed to be $\approx10^5$ and the respective QCRB is observed to be $\approx14470$, On increasing the event numbers we observe the same pattern for the uncertainty: it linearly decreases. For $10^7$ events, the CCRB drops to $\approx60$ and the respective QCRB drops to $\approx 5.38$ , this shows the enormous drop in uncertainty.
\end{itemize}

We observe that, in its current form, KM3NeT provides stronger constraints on the new-physics parameter, as reflected by the CRB-derived uncertainty bounds. Nevertheless, improved measurements of the muon-neutrino appearance probability would benefit from larger event statistics. In contrast, while increasing the number of events in IceCube leads to a reduction in the uncertainty on $\epsilon_{ss}$.
\vspace{1cm}

For $\lambda = m_{s}$,
\begin{figure}[H]
  \centering
  \includegraphics[width=0.9\columnwidth]{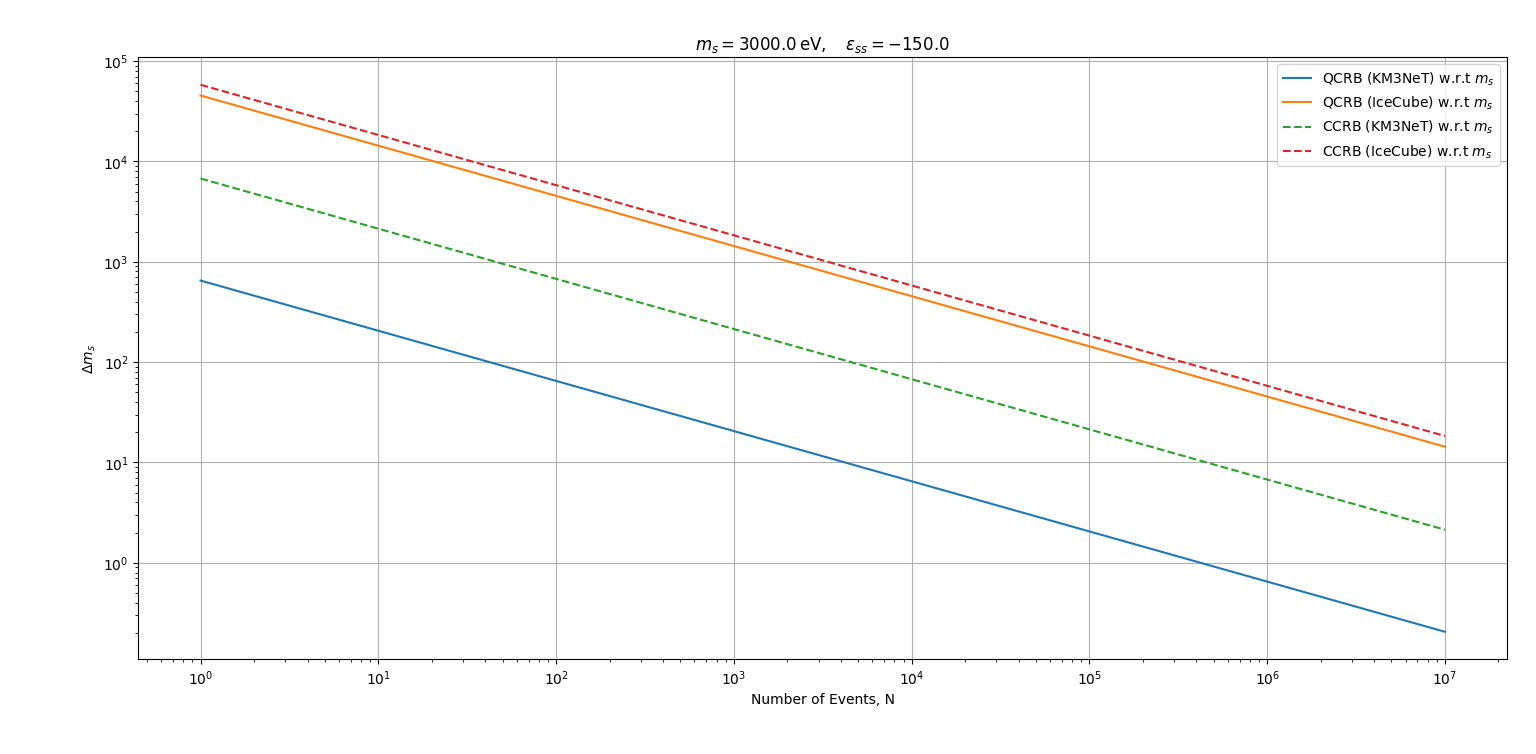}
  \caption{$\Delta m_s$ vs.\ No. of events $(N)$ for
    \textbf{KM3NeT and IceCube} (MSW scenario).
    Parameters as in Fig.~\ref{fig:FQ_KM3_MSW_ess}.
    The absolute scale is smaller than for $\ess$ by
    several orders of magnitude, reflecting the reduced
    sensitivity of $\Psmu$ to small variations in $\ms$
    at fixed energy.}
  \label{fig:CRB_MSW_ms}
\end{figure}

For $\lambda = m_s$,

For KM3NeT,
\begin{itemize}
    \item For the current scenario (i.e., $N=1$), the CCRB is observed to be $\approx 6983.6$ and the respective QCRB is observed to be $\approx 663.2$. For $10^{7}$ events, the CCRB drops to $\approx 2.2$ and the respective QCRB drops to $\approx 0.21$, this shows the enormous drop in uncertainty.
\end{itemize}

For IceCube,
\begin{itemize}
    \item For the current scenario (i.e., $N=1$), the CCRB is observed to be $\approx 58921.4$ and the respective QCRB is observed to be $\approx 44667.2$. For $10^{7}$ events, the CCRB drops to $\approx 19.6$ and the respective QCRB drops to $\approx 14.51$, this shows the enormous drop in uncertainty.
\end{itemize}

We observe that, as with $\varepsilon_{ss}$, KM3NeT provides substantially tighter constraints on $m_s$ than IceCube at every value of $N$, with the CCRB smaller by roughly an order of magnitude ($\approx8.4\times$) and the QCRB smaller by nearly two orders of magnitude ($\approx67\times$) at KM3NeT's baseline. The quantum advantage, quantified by the ratio $\Delta_{\mathrm{CCRB}}/\Delta_{\mathrm{QCRB}}$, is $\approx 10.5$ for KM3NeT compared to only $\approx 1.3$--$1.4$ for IceCube, and this ratio remains essentially unchanged as $N$ increases from $1$ to $10^{7}$. This indicates that the disparity in quantum-optimal sensitivity between the two detectors is an intrinsic, geometry-driven feature of the propagating $|\nu_s\rangle$ state rather than a statistical effect that can be removed by accumulating more events. As with $\varepsilon_{ss}$, increasing the number of events in IceCube does reduce the uncertainty on $m_s$, but even at $N=10^{7}$ IceCube's bounds remain looser than KM3NeT's bounds at $N=1$, underscoring that KM3NeT's baseline placement, rather than event statistics alone, is primarily responsible for its superior sensitivity to the sterile-neutrino mass parameter.

\subsection{CRB for NSI scenario}\label{sec:value_of_QFI_NSI}
\begin{figure}[H]
  \centering
  \includegraphics[width=0.9\columnwidth]{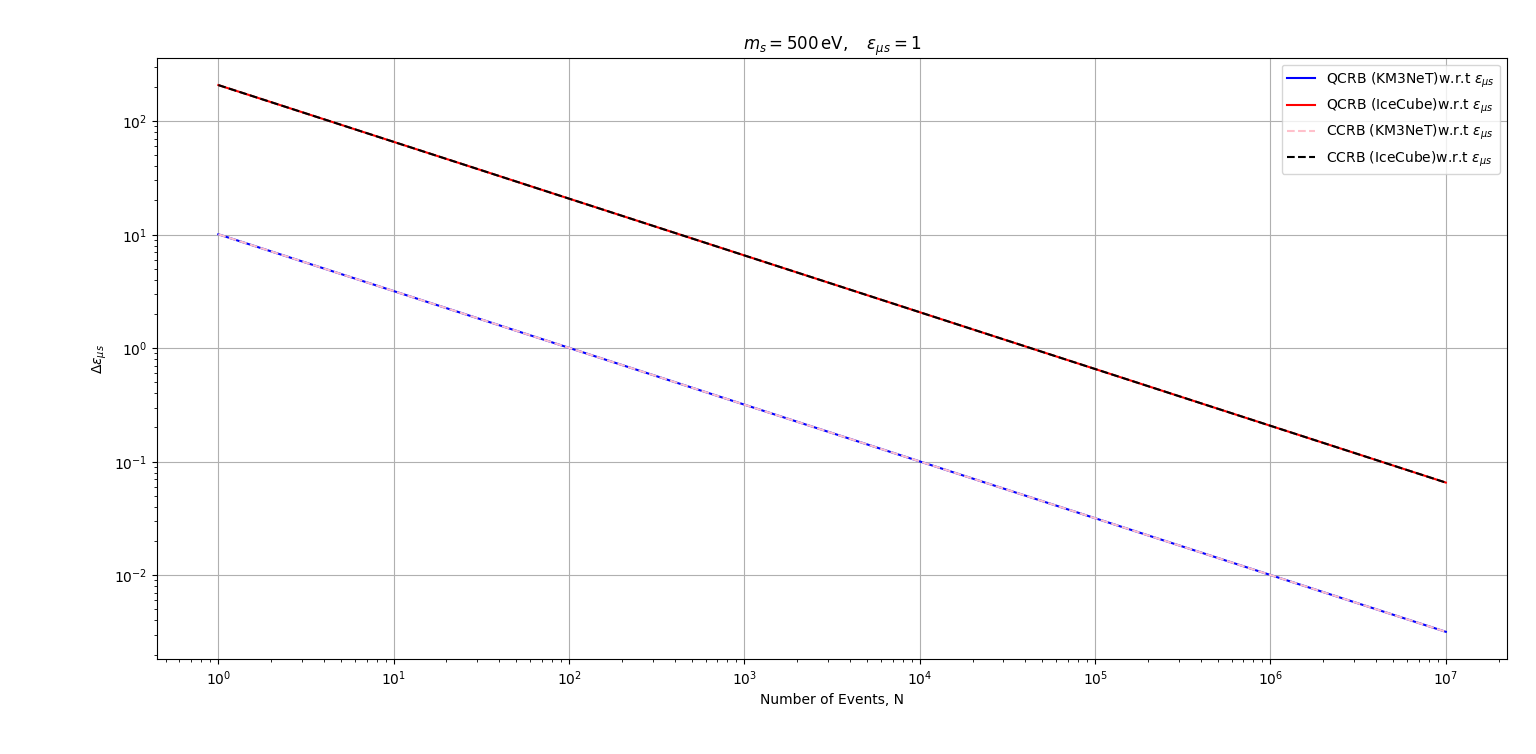}
  \caption{$\Delta \epsilon_{\mu s}$ vs.\ No. of events $(N)$ for
    \textbf{KM3NeT and IceCube} (NSI scenario).
    Parameters as in Fig.~\ref{fig:FQ_KM3_NSI_ems}.
    The absolute scale is smaller than for $\ems$ by
    several orders of magnitude, reflecting the reduced
    sensitivity of $\Psmu$ to small variations in $\ms$
    at fixed energy.}
  \label{fig:CRB_NSI_eps_mus}
\end{figure}

It is evident from the plots that as the events capture numbers increase, the uncertainty associated with the parameters steadily decreases. Talking about Figure \ref{fig:CRB_NSI_eps_mus}, we have set the baseline length to KM3NeT ($\approx 147$ km) and IceCube ($\approx 14$ km) and carried out our study (QCRB and CCRB) by varying the Number of events (N) for both the experiments. \\

For $\lambda = \ems$,

For KM3NeT,
\begin{itemize}
    \item For the current scenario (i.e, N = 1), the CCRB = QCRB and is observed to be $\approx10.11$, this ensures that as these observations have been encoded with such large uncertainties so for more accurate analysis, the event number must be increased: On increasing the event numbers, we observe that the uncertainty linearly decreases. For $10^7$ events, CCRB = QCRB drops to $\approx3.29 \times 10^{-3}$, which shows there is an enormous drop in uncertainty.
\end{itemize}
For IceCube,
\begin{itemize}
    \item For the current scenario (i.e, N = 1), the CCRB = QCRB and is observed to be $\approx209.9$, this ensures that as these observations have been encoded with such large uncertainties so for more accurate analysis, the event number must be increased: On increasing the event numbers, we observe that the uncertainty linearly decreases. For $10^7$ events, CCRB = QCRB drops to $\approx6.62 \times 10^{-2}$, which shows there is an enormous drop in uncertainty.
\end{itemize}
We observe that, in its current form, KM3NeT provides stronger constraints on the new-physics parameter, as reflected by the CRB-derived uncertainty bounds. Nevertheless, improved measurements of the muon-neutrino appearance probability would benefit from larger event statistics. In contrast, while increasing the number of events in IceCube leads to a reduction in the uncertainty on $\ems$.
Since $\FC=\FQ$ in the NSI scenario, this means that there is no information loss and 100 \% of the information is extracted about the $\lambda=\ems$ parameter. This leads to the condition of CCRB=QCRB, as shown in Fig. \ref{fig:CRB_NSI_eps_mus}.\\ \newline
For $\lambda = m_s$,
\begin{figure}[H]
  \centering
  \includegraphics[width=0.9\columnwidth]{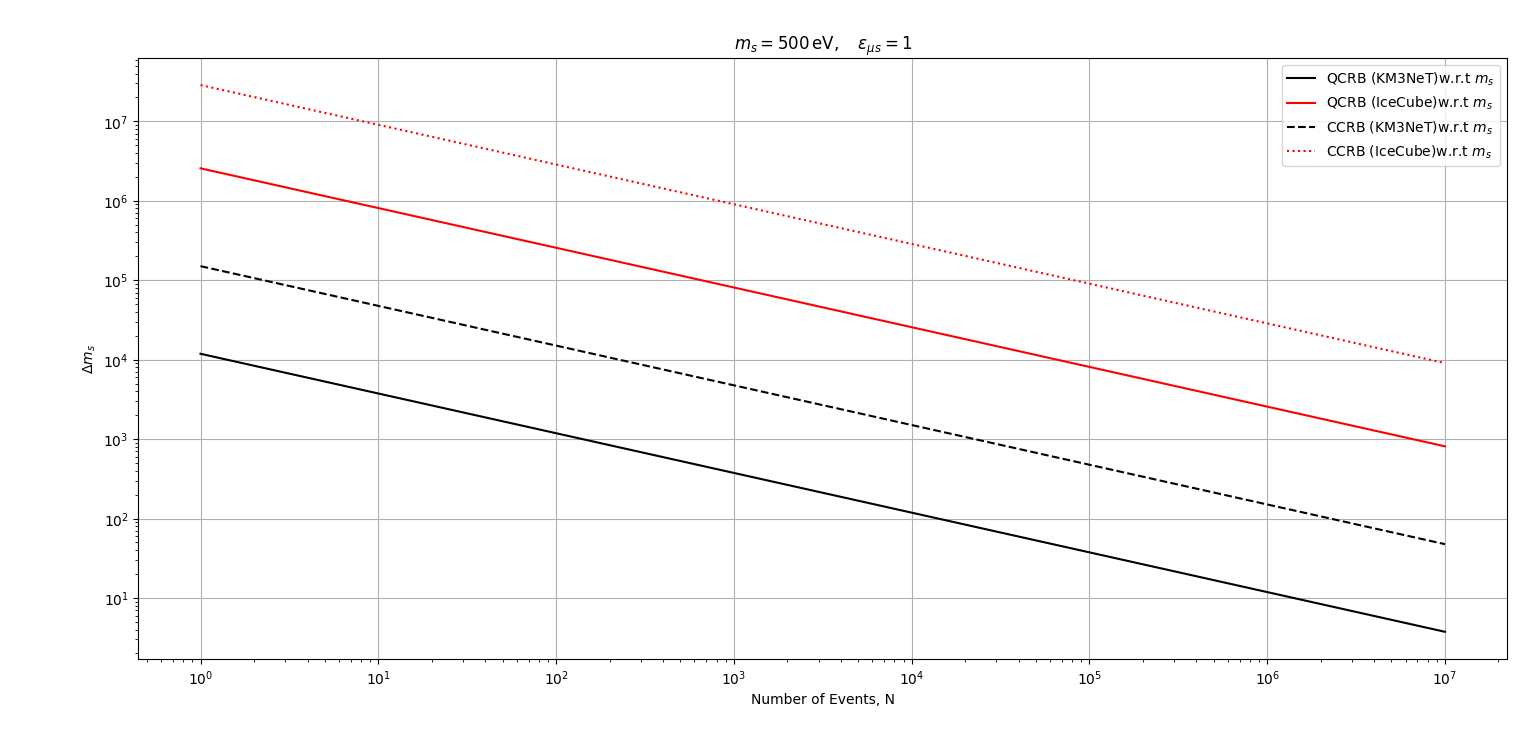}
  \caption{$\Delta m_s$ vs.\ No. of events $(N)$ for
    \textbf{KM3NeT and IceCube} (NSI scenario).
    Parameters as in Fig.~\ref{fig:FQ_KM3_NSI_ems}.
    The absolute scale is smaller than for $\ems$ by
    several orders of magnitude, reflecting the reduced
    sensitivity of $\Psmu$ to small variations in $\ms$
    at fixed energy.}
  \label{fig:CRB_NSI_ms}
\end{figure}

For KM3NeT,
\begin{itemize}
    \item For the current scenario (i.e., $N=1$), the CCRB is observed to be $\approx 151012$ and the respective QCRB is observed to be $\approx 11246.4$. For $10^{7}$ events, the CCRB drops to $\approx 47.22$ and the respective QCRB drops to $\approx 3.77$, this shows the enormous drop in uncertainty.
\end{itemize}

For IceCube,
\begin{itemize}
    \item For the current scenario (i.e., $N=1$), the CCRB is observed to be $\approx 2.98 \times 10^{7}$ and the respective QCRB is observed to be $\approx 2.61 \times 10^{6}$. For $10^{7}$ events, the CCRB drops to $\approx 9.12793 \times 10^{3}$ and the respective QCRB drops to $\approx 837.56$, this shows the enormous drop in uncertainty.
\end{itemize}
We observe that, as with $\ems$, KM3NeT provides substantially tighter constraints on $m_s$ than IceCube at every value of $N$, with both the CCRB and QCRB smaller by roughly 2 orders of magnitude at KM3NeT's baseline. The quantum advantage, quantified by the ratio $\Delta_{\mathrm{CCRB}}/\Delta_{\mathrm{QCRB}}$, is $\approx 13.4 $ for KM3NeT compared to $\approx 11.42$ for IceCube, and this ratio remains essentially unchanged as $N$ increases from $1$ to $10^{7}$, where $\Delta_{\mathrm{CCRB}}/\Delta_{\mathrm{QCRB}}$ is $\approx 12.5$ for Km3Net and is $\approx 10.9$ for IceCube. This indicates that the disparity in quantum-optimal sensitivity between the two detectors is an intrinsic, geometry-driven feature of the propagating $|\nu_s\rangle$ state rather than a statistical effect that can be removed by accumulating more events. As with $\ems$, increasing the number of events in IceCube does reduce the uncertainty on $m_s$, but even at $N=10^{7}$ IceCube's bounds remain looser than KM3NeT's bounds at $N=1$, underscoring that KM3NeT's baseline placement, rather than event statistics alone, is primarily responsible for its superior sensitivity to the sterile-neutrino mass parameter.

\section{Optimal Baseline for IceCube and KM3NeT}
We have studied the estimation limits — quantum and classical Fisher information (QFI and CFI) — for both KM3NeT and IceCube from the MSW and NSI perspectives. In this section, we retain the same benchmark potentials as in these experiments while varying the baseline length, in order to identify the regions where the QFI and CFI become most sensitive. This analysis provides useful guidance for the design of future experimental setups focusing on the choice of an optimal baseline length.

\begin{figure}[H]
    \centering
    \begin{subfigure}[b]{0.45\textwidth}
        \centering
        \includegraphics[width=\textwidth]{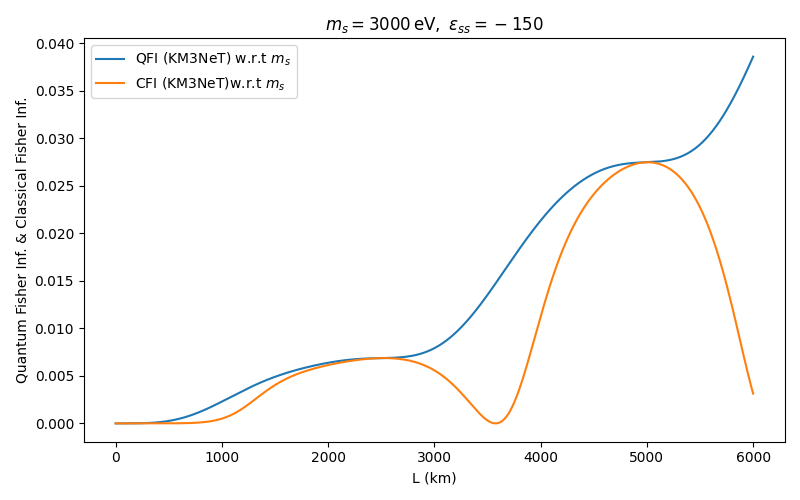}
        \caption{MSW Effect on QFI and CFI by varying $m_s$ for KM3NeT}
        \label{fig:QFI_KM3_MSW_ms_opt}
    \end{subfigure}
    \hfill
    \begin{subfigure}[b]{0.45\textwidth}
        \centering
        \includegraphics[width=\textwidth]{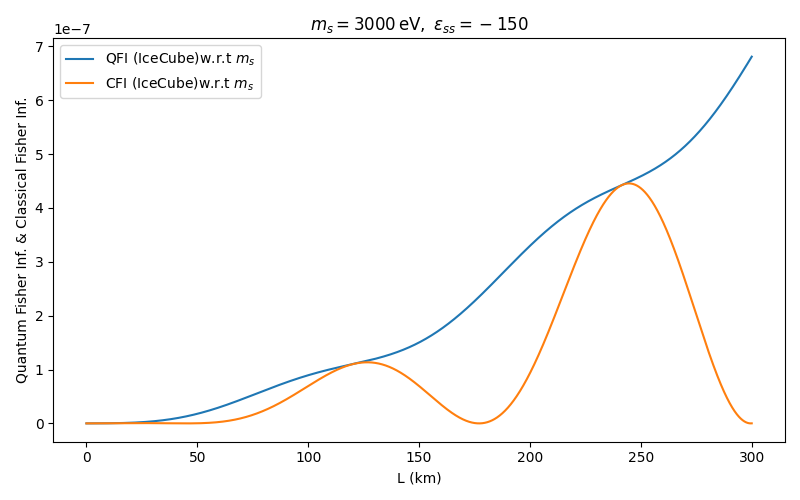}
        \caption{MSW Effect on QFI and CFI by varying $m_s$ for IceCube}
        \label{fig:QFI_IC_MSW_ms_opt}
    \end{subfigure}
    \caption{Analysis of $m_s$ for Km3NeT and IceCube for MSW effect with extended baseline}
    \label{fig:main}
\end{figure}

From Fig.~13a and Fig.~13b, we study how the QFI and CFI with respect to $m_s$ evolve for
KM3NeT and IceCube as the baseline length $L$ is extended well beyond the physical detector
baselines, allowing us to identify the regions of $L$ where sensitivity to the sterile-neutrino
mass is maximized. \\

For KM3NeT, the QFI and CFI with respect to $m_s$ range over $[0, 0.04]$ for baselines
$L \in [0, 6000]\,\mathrm{km}$ (Fig.~13a). For IceCube, a qualitatively similar peak
structure is observed, though over a much shorter baseline range and smaller absolute
scale: the QFI and CFI range over $[0, 7 \times 10^{-7}]$ for $L \in [0, 300]\,\mathrm{km}$
(Fig.~13b). For both KM3NeT and IceCube, the QFI and CFI are observed to follow a periodic
structure in $L$. Focusing on the QFI, we observe that it increases monotonically, briefly
saturates, and then rises again, reflecting a growing information content about $m_s$
carried by the propagating $|\nu_s\rangle$ state as the baseline is extended. The CFI, by
contrast, peaks periodically at an interval of $\sim 120\,\mathrm{km}$, touching the QFI
envelope at each peak before falling away sharply at the intervening oscillation nodes.

This periodic coincidence of $F_Q$ and $F_C$ is physically significant: at these baselines,
the flavor-projection measurement already performed by neutrino telescopes saturates the
quantum bound, so no alternative measurement strategy could extract more information about
$m_s$ than is already accessible via standard muon-track detection. This suggests that an
IceCube-like detector, if extended to a baseline near a multiple of $\sim 120\,\mathrm{km}$,
would be positioned at a point of maximal estimation precision for $m_s$, since $F_Q = F_C$
there ensures the standard measurement already achieves quantum-optimal sensitivity.
Consequently, we argue that future or upgraded neutrino telescope geometries should be
designed, where possible, to fall near these $F_Q = F_C$ baselines, as doing so guarantees
that ordinary flavor-counting analyses are not leaving precision on the table relative to
the fundamental quantum limit.

\begin{figure}[H]
    \centering
    \begin{subfigure}[b]{0.45\textwidth}
        \centering
        \includegraphics[width=\textwidth]{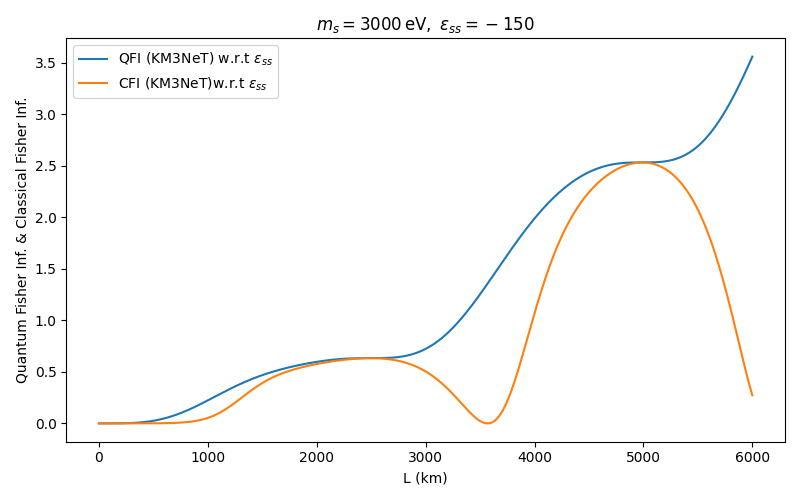}
        \caption{MSW Effect on QFI and CFI by varying $\ess$ for KM3NeT}
        \label{fig:QFI_KM3_MSW_eps_ss_opt}
    \end{subfigure}
    \hfill
    \begin{subfigure}[b]{0.45\textwidth}
        \centering
        \includegraphics[width=\textwidth]{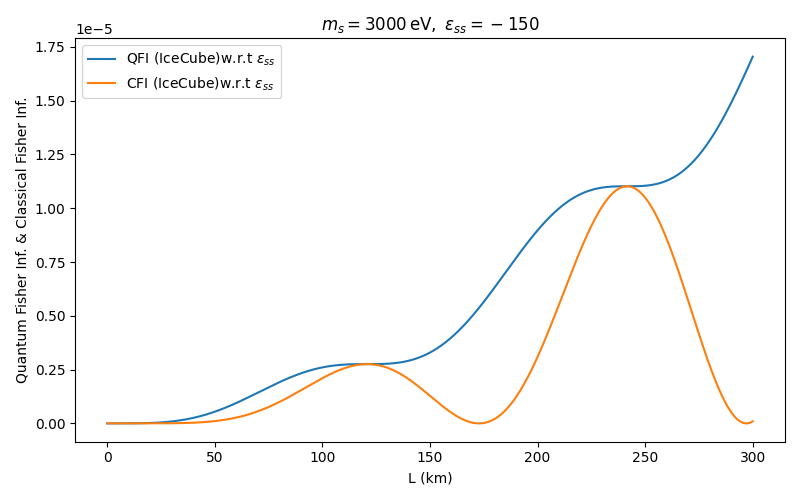}
        \caption{MSW Effect on QFI and CFI by varying $\ess$ for IceCube}
        \label{fig:QFI_IC_MSW_eps_ss_opt}
    \end{subfigure}
    \caption{Analysis of $\ess$ for Km3NeT and IceCube for MSW effect with extended baseline}
    \label{fig:main}
\end{figure}
For KM3NeT, the QFI and CFI with respect to $\varepsilon_{ss}$ range over $[0, 3.5]$ for
baselines $L \in [0, 6000]\,\mathrm{km}$ (Fig.~14a). For IceCube, a qualitatively similar peak
pattern is observed, but the QFI and CFI values range over $[0, 1.75 \times 10^{-5}]$ for
respective baseline $L \in [0, 300]\,\mathrm{km}$ (Fig.~14b). For both KM3NeT and IceCube, it
is clearly observed that the QFI and CFI follow a periodic cycle. Focusing on the QFI, we
observe that it keeps on increasing and then saturates, meaning that the information about the
sterile neutrino state (w.r.t $\varepsilon_{ss}$) increases and then gets saturated, and again
increases; and the CFI basically peaks up periodically at a range of 120 km gap. We observe the
peaks meaning that if we make an experiment like IceCube extending its baseline for a cycle of
120 km, we can have a case where QFI = CFI leading to maximum estimation of the parameter
$\varepsilon_{ss}$.

Taken together, the extended-baseline analysis for both $m_s$ and $\varepsilon_{ss}$ suggests
that the two detectors admit different practically optimal baselines. For KM3NeT, the QFI and
CFI curves coincide over a broad plateau spanning $L \sim 2100$–$2600\,\mathrm{km}$, indicating
that an extension to this range would allow KM3NeT to approach quantum-optimal sensitivity;
however, physically realizing such a baseline for KM3NeT is far from trivial, since it would
require a fundamentally different source-detector geometry than is currently available. For
IceCube, by contrast, the $F_Q = F_C$ condition is already satisfied at the much more modest
baseline of $L \approx 120\,\mathrm{km}$, which is a comparatively minor extension given
IceCube's already-short baseline of $14\,\mathrm{km}$. Notably, the QFI and CFI achieved by
IceCube at this extended $120\,\mathrm{km}$ baseline approach the same order as those already
achieved by KM3NeT at its current $147\,\mathrm{km}$ baseline ($F_Q \approx 2.6\times10^{-4}$,
$F_C \approx 3.5\times10^{-7}$ for $\varepsilon_{ss}$), and fall within the same order of
magnitude for $m_s$. This indicates that a comparatively modest baseline extension for IceCube
would be sufficient to bring its sensitivity to $\{\varepsilon_{ss}, m_s\}$ close to that
currently enjoyed by KM3NeT, while simultaneously guaranteeing $F_Q = F_C$ saturation at that
baseline. We therefore argue that extending IceCube's effective baseline to $\sim 120\,\mathrm{km}$
is a more practically achievable route to improved new-physics sensitivity than attempting to
extend KM3NeT to $\sim 2500\,\mathrm{km}$, and could allow IceCube to provide an estimation of
$\varepsilon_{ss}$ and $m_s$ that is both quantum-optimal and competitive with KM3NeT's current
reach.
\begin{figure}[H]
    \centering
    \begin{subfigure}[b]{0.45\textwidth}
        \centering
        \includegraphics[width=\textwidth]{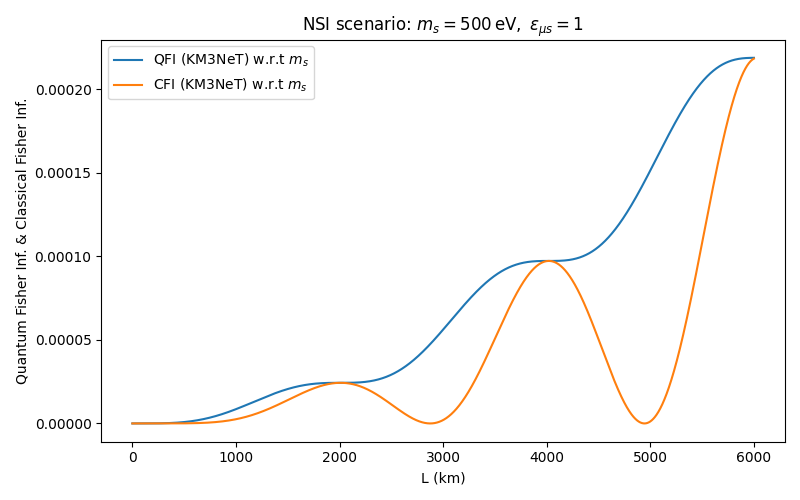}
        \caption{NSI Effect on QFI and CFI by varying $m_s$ for Km3NeT}
        \label{fig:QFI_KM3_NSI_ms_opt}
    \end{subfigure}
    \hfill
    \begin{subfigure}[b]{0.45\textwidth}
        \centering
        \includegraphics[width=\textwidth]{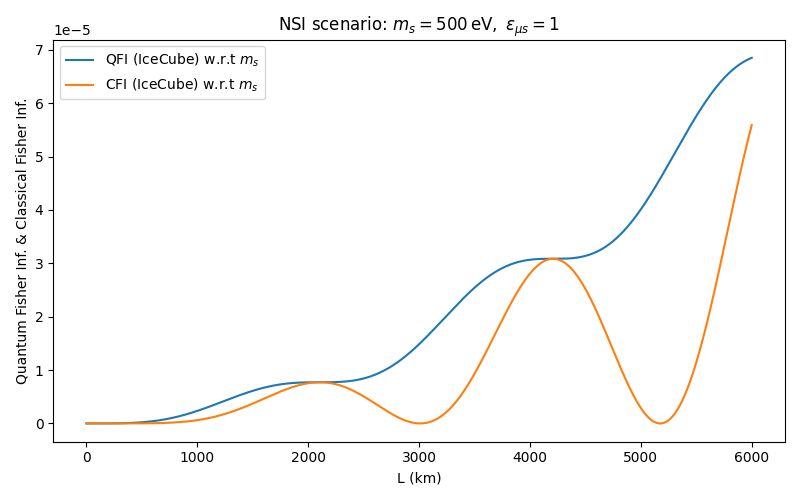}
        \caption{NSI Effect on QFI and CFI by varying $m_s$ for IceCube}
        \label{fig:QFI_IC_NSI_ms_opt}
    \end{subfigure}
    \caption{Analysis of $m_s$ for Km3NeT and IceCube for NSI effect with extended baseline}
    \label{QFI_NSI_m_s_ext_baseline}
\end{figure}
Figs.~\ref{fig:QFI_KM3_NSI_ms_opt} and \ref{fig:QFI_IC_NSI_ms_opt} extend the
baseline to $L \in [0, 6000]\,\mathrm{km}$, well beyond the geometries of current
detectors, for the NSI scenario. We present the QFI and CFI for the parameter
$m_s$ for both KM3NeT and IceCube over this extended range, with the aim of
identifying the baseline regions at which sensitivity to the sterile-neutrino mass
is maximized. The Fisher information ranges over $[0,\, 2 \times 10^{-4}]$ for
KM3NeT and $[0,\, 7 \times 10^{-5}]$ for IceCube, reflecting $\approx$
one order of magnitude difference in both the QFI and CFI between the two
detectors across all baselines considered. The QFI in both cases displays a
periodic envelope, increasing monotonically within each period before saturating,
with successive maxima occurring at intervals of approximately $L \approx 2000$~km.
The CFI exhibits a complementary periodic structure, reaching sharp peaks at the
same baseline intervals at which it touches the QFI envelope. Notably, for KM3NeT
the CFI approaches the QFI envelope near $L \approx 6000$~km, whereas it is not observed for IceCube within the baseline range considered. At these
special baselines, the saturation condition $F_C = F_Q$ is observed, implying
zero information loss and quantum-optimal sensitivity to $m_s$. These results
suggest that if future neutrino telescopes with baselines extending to
$L \approx 2000$~km were to become feasible, the NSI scenario would admit maximum
estimation precision for $m_s$, with the classical flavor-projection measurement
saturating the fundamental quantum limit.
\begin{figure}[H]
    \centering
    \begin{subfigure}[b]{0.45\textwidth}
        \centering
        \includegraphics[width=\textwidth]{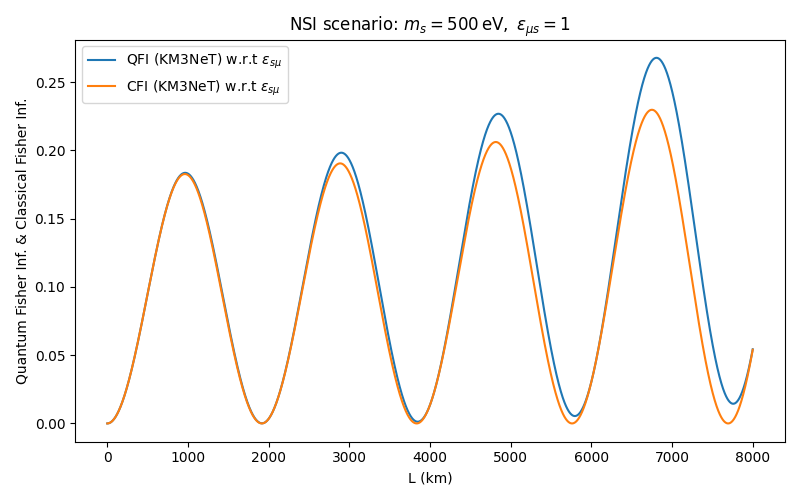}
        \caption{NSI Effect on QFI and CFI by varying $\ems$ for KM3NeT}
        \label{fig:QFI_KM3_NSI_eps_opt}
    \end{subfigure}
    \hfill
    \begin{subfigure}[b]{0.45\textwidth}
        \centering
        \includegraphics[width=\textwidth]{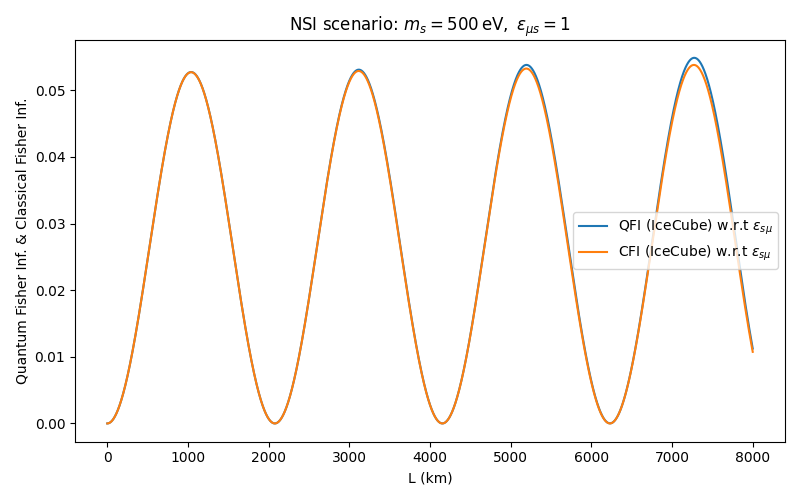}
        \caption{NSI Effect on QFI and CFI by varying $\ems$ for IceCube}
        \label{fig:QFI_IC_NSI_eps_opt}
    \end{subfigure}
    \caption{Analysis of $\ems$ for Km3NeT and IceCube for NSI effect with extended baseline}
    \label{QFI_NSI_eps_mus_ext_baseline}
\end{figure}

For KM3NeT, the QFI and CFI with respect to $\varepsilon_{\mu s}$ oscillate periodically with
baseline, with peak amplitudes growing from $\approx 0.18$ near $L\approx1000\,\mathrm{km}$ to
$\approx 0.27$ near $L\approx7000\,\mathrm{km}$ (Fig.~16a), consistent with the $L^2$-growing
envelope characteristic of the NSI conversion probability. For IceCube, a qualitatively similar
oscillatory pattern is observed over the same extended baseline range, though the peak amplitude
remains comparatively flat, saturating around $\approx 0.05$ across successive peaks
(Fig.~16b). Unlike the MSW scenario, where the CFI is punctuated by sharp dips at oscillation
nodes while the QFI stays high, here the QFI and CFI curves are numerically indistinguishable
at essentially every baseline, for both KM3NeT and IceCube. This confirms, over the full
extended-baseline range, the exact saturation $F_Q = F_C$ established analytically in
Sec.~3.4.1 for the NSI $\varepsilon_{\mu s}$ parameter: because the underlying pure-state
condition and binary flavor-projection measurement together force the SLD to be diagonal in
the flavor basis regardless of $L$, the standard muon-track measurement remains quantum-optimal
at every baseline, not merely at isolated crossing points as in the MSW case.

The successive peaks recur with a spacing of $\sim 1800$–$2000\,\mathrm{km}$ for both detectors,
and the growing envelope indicates that sensitivity to $\varepsilon_{\mu s}$ continues to improve
with baseline rather than saturating, in contrast to the MSW parameters. This is consistent with
the optimal baseline of $L_{\mathrm{opt}} \sim 1300\,\mathrm{km}$ identified for the NSI scenario,
where the conversion probability envelope is maximized; since $F_Q=F_C$ holds at all baselines
in this scenario, any detector positioned near this optimal window would automatically achieve
quantum-optimal sensitivity to $\varepsilon_{\mu s}$ using only conventional flavor-counting
analysis, without requiring any exotic measurement strategy.

\section{The Chi-Square Formalism and the Advantage of Fisher Information Analysis}
\label{sec:chi2}

\subsection{The \texorpdfstring{$\chi^2$}{chi-square} formalism}
\label{sec:chi2_formalism}

In standard neutrino oscillation analyses, sensitivity to a new-physics
parameter $\lambda$ (here $\lambda \in \{\ess, \ems, \ms\}$) is assessed
through a $\chi^2$ statistic constructed from the observed and predicted
event counts. For a single measurement channel with expected number of
muon-neutrino events $N_{\rm exp}(\lambda)$ and observed count $N_{\rm obs}$,
the simplest Gaussian form is
\begin{equation}
    \chi^2(\lambda) = \frac{\left[N_{\rm obs} - N_{\rm exp}(\lambda)\right]^2}
    {\sigma^2(\lambda)},
    \label{eq:chi2_gauss}
\end{equation}
while in the Poisson regime relevant to the small event samples
characteristic of ultra-high-energy neutrino telescopes,
\begin{equation}
    \chi^2(\lambda) = 2\left[N_{\rm exp}(\lambda) - N_{\rm obs}
    + N_{\rm obs}\ln\!\frac{N_{\rm obs}}{N_{\rm exp}(\lambda)}\right].
    \label{eq:chi2_poisson}
\end{equation}
A confidence interval on $\lambda$ is obtained from the standard
$\Delta\chi^2 = \chi^2(\lambda) - \chi^2(\lambda_{\rm true})$ criterion, with
$\Delta\chi^2 = 1$ conventionally defining the $1\sigma$ region for a single
parameter.

This construction is not an independent alternative to the Fisher
information framework used throughout this paper. Expanding $\chi^2(\lambda)$
to second order about its minimum,
\begin{equation}
    \chi^2(\lambda) \approx \chi^2_{\rm min}
    + \frac{1}{2}\left.\frac{\partial^2 \chi^2}{\partial \lambda^2}
    \right|_{\lambda_{\rm true}} (\lambda - \lambda_{\rm true})^2,
    \label{eq:chi2_expansion}
\end{equation}
and comparing with the definition of $\FC$ in Eqs.~\eqref{eq:CFI_binary},
the curvature of the $\chi^2$ surface at its minimum is asymptotically
identical to the classical Fisher information:
\begin{equation}
    \left.\frac{\partial^2 \chi^2}{\partial \lambda^2}\right|_{\lambda_{\rm true}}
    = 2\,\FC(\lambda_{\rm true}).
    \label{eq:chi2_FI_relation}
\end{equation}
Consequently, the $1\sigma$ interval from $\Delta\chi^2=1$,
$\Delta\lambda_{\chi^2} = 1/\sqrt{\FC(\lambda)}$, coincides exactly with the
classical Cram\'er--Rao bound of Eq.~\eqref{eq:CCRB_def}. A conventional
$\chi^2$ fit performed by a neutrino telescope collaboration is therefore,
under the Gaussian/Poisson approximations above, a numerical estimate of the
same classical Fisher information computed directly in
Sec.~\ref{sec:QFI_MSW}--\ref{sec:QFI_NSI}, obtained via a specific test
statistic, binning scheme, and treatment of systematics rather than via the
analytic derivative $\partial_\lambda\Psmu$. The two methods, correctly
implemented, must therefore agree in the asymptotic limit; we note that this
asymptotic equivalence is itself expected to break down precisely in the
$N=1$ regime relevant to the single KM3-230213A event, which is one reason
the event-count scaling of Eq.~\eqref{eq:Nevents} below is informative in
its own right.

What the $\chi^2$ formalism cannot do, by construction, is address whether
the measurement itself --- flavor projection via muon-track identification
--- is the optimal one available. A $\chi^2$ analysis presupposes a fixed
measurement scheme and asks only how well $\lambda$ can be constrained given
that scheme; it has no notion of $\FQ(\lambda)$ and therefore no way of
establishing whether $\FC(\lambda) = \FQ(\lambda)$, or whether some
alternative measurement could extract more information from the same
propagating $\ket{\nu_s}$ state. This is precisely the question addressed in
Secs.~\ref{sec:QFI_MSW}--\ref{sec:QFI_NSI}.

\subsection{Advantage of Fisher information analysis over \texorpdfstring{$\chi^2$}{chi-square} analysis}
\label{sec:advantage}

Building on the equivalence established above, the QFI framework provides
the following, understood as extensions beyond what a $\chi^2$ analysis is
designed to answer, not as a claim that conventional $\chi^2$ fits are
performed incorrectly:

\begin{enumerate}

  \item \textbf{Measurement-independence.}
    The QCRB~\eqref{eq:QCRB_def} holds for \emph{any} unbiased measurement
    strategy --- including future exotic ones (phase-sensitive detectors,
    quantum state tomography). We showed in Sec.~\ref{sec:QFI_NSI} that the
    existing flavor-projection measurement already saturates it exactly for
    $\ems$ at every baseline, and periodically for $\ess$ and $\ms$ in the
    MSW scenario (Sec.~\ref{sec:QFI_MSW}); this closes the search for better
    strategies in those regimes.

  \item \textbf{Baseline optimization.}
    By computing $\FQ(\lambda;L)$ as a continuous function of $L$
    (Sec.~5), we identified the optimal baseline windows summarized in
    Fig.~\ref{fig:CRB_MSW_eps_ss}. A conventional $\chi^2$ analysis only
    evaluates sensitivity at the detector baselines that currently exist.

  \item \textbf{Information-theoretic interpretation.}
    The ratio $\FQ^{\KM}/\FQ^{\IC}$ quantifies a fundamental asymmetry in
    quantum information content, not a statistical fluctuation. For
    $\ess$ at the present baselines, $\FQ^{\KM}/\FQ^{\IC}
    \approx (2.6\times10^{-4})/(4.5\times10^{-9}) \approx 5.8\times10^4$,
    demonstrating that the KM3NeT--IceCube tension has an
    information-theoretic component independent of luminosity or detector
    efficiency.

  \item \textbf{Event-count scaling.}
    The QCRB gives the minimum number of IceCube events needed to match
    KM3NeT's single-event precision on $\lambda$. For the MSW scenario,
    since $\QCRB \propto 1/\sqrt{N\,\FQ}$, equating the two bounds at
    matching precision requires
    \begin{align}
      N_{\IC}\,\FQ^{\IC} &= N_{\KM}\,\FQ^{\KM}
      \nonumber\\
      \Rightarrow\quad N_{\IC} &= N_{\KM}\,\frac{\FQ^{\KM}}{\FQ^{\IC}}
        \approx \left(\frac{\QCRB^{\IC}(\ess;N{=}1)}
        {\QCRB^{\KM}(\ess;N{=}1)}\right)^2
        \approx \left(\frac{14470}{61}\right)^2
        \approx 5.6\times10^{4},
      \label{eq:Nevents}
    \end{align}
    using the $N=1$ values reported in Sec.~\ref{sec:value_of_QFI}. This is
    consistent, to the expected rounding precision, with the direct
    $\FQ^{\KM}/\FQ^{\IC}$ ratio above. An uncertainty ratio of
    $\approx 237$ on $\ess$ between the two detectors therefore implies an
    event-count ratio of $\approx 237^2 \approx \mathcal{O}(10^4$--$10^5)$
    --- IceCube would need on the order of $5\times10^4$ comparable
    ultra-high-energy events to match KM3NeT's present single-event
    precision on $\ess$, underscoring how substantial an observational
    program would be required to close this gap through statistics alone.
\end{enumerate}
\section{Conclusions}
\label{sec:conclusions}

We have applied Quantum Fisher Information (QFI) and Classical Fisher
Information (CFI) to the two sterile-neutrino scenarios --- MSW resonance
and non-standard interactions (NSI) --- proposed by Brdar and Chattopadhyay
\cite{Brdar:2026} to explain the tension between the KM3-230213A event and
IceCube's non-observation of a comparable signal. Our analysis yields four
main results.

First, for the \textbf{NSI scenario}, we proved analytically
(Sec.~\ref{sec:QFI_NSI}) that $\FQ = \FC$ for $\ems$ at every baseline $L$,
for both KM3NeT ($\FQ=\FC\approx9.46\times10^{-3}$) and IceCube
($\FQ=\FC\approx2.05\times10^{-5}$). This follows directly from the
pure-state character of the propagating $\ket{\nu_s}$ together with the
binary flavor-projection measurement, which together force the symmetric
logarithmic derivative to be diagonal in the flavor basis regardless of
baseline. Standard muon-track detection, exactly as performed by existing
neutrino telescopes, is therefore already quantum-optimal for $\ems$; no
alternative measurement strategy could extract additional information about
this parameter from the propagating state.

Second, for the \textbf{MSW scenario}, we found $\FQ \geq \FC$ strictly at
the current detector baselines, with KM3NeT losing $\approx 3$ orders of
magnitude of the available quantum information about $\ess$
($\FQ\approx2.6\times10^{-4}$ vs.\ $\FC\approx3.5\times10^{-7}$) and IceCube
losing $\approx 2$ orders ($\FQ\approx4.5\times10^{-9}$ vs.\
$\FC\approx7.4\times10^{-11}$). For $\ms$, both detectors lose
$\approx 2$ orders of magnitude at their present baselines (see
Sec.~\ref{sec:m_s_MSW}); saturation $\FQ=\FC$ is recovered only
periodically, at isolated baselines spaced by $\sim120\,\mathrm{km}$,
reflecting the resonant character of sterile-to-active conversion in this
scenario as opposed to the monotonic $L^2$-growth of the NSI case. For the
NSI scenario, both detectors lose $\approx2$ orders of magnitude of
information about $\ms$ at their current baselines
(Sec.~\ref{sec:m_s_NSI}), in contrast to the exact $\ems$ saturation noted
above.

Third, using the corresponding Cram\'er--Rao bounds (Sec.~\ref{sec:CRB}), we
showed that KM3NeT constrains $\ess$, $\ems$, and $\ms$ more tightly than
IceCube by roughly one to two orders of magnitude at $N=1$, and that the
ratio $\Delta_{\rm CCRB}/\Delta_{\rm QCRB}$ --- the quantum advantage ---
remains essentially fixed as $N$ is increased from $1$ to $10^7$ simulated
events (e.g.\ $\approx10.5$ for KM3NeT vs.\ $\approx1.3$--$1.4$ for IceCube
in the case of $\ms$/MSW; $\approx13.4\to12.5$ for KM3NeT vs.\
$\approx11.4\to10.9$ for IceCube in the case of $\ms$/NSI). This
demonstrates that the KM3NeT--IceCube sensitivity asymmetry is an intrinsic,
geometry-driven property of the propagating sterile-neutrino state, encoded
in the baseline-dependent Fisher information itself, rather than a
statistical artifact that additional exposure time alone could remove.
Equation~\eqref{eq:Nevents} quantifies this concretely: bringing IceCube to
parity with KM3NeT's present single-event precision on $\ess$ would require
accumulating $\mathcal{O}(10^4\text{--}10^5)$ comparable high-energy events.

Fourth, extending the analysis to baselines beyond current detector
geometries (Sec.~5), we identified specific windows --- $L\approx120\,
\mathrm{km}$ for an IceCube-like detector and $L\sim2100$--$2600\,
\mathrm{km}$ for a KM3NeT-like detector in the MSW scenario, and
$L_{\rm opt}\sim1300\,\mathrm{km}$ ($\ems$) to $\sim2000\,\mathrm{km}$
($\ms$) in the NSI scenario --- at which $\FQ=\FC$ is recovered or the
conversion-probability envelope is maximized. We stress that these
optimal-baseline estimates are necessarily \emph{conditional} on the present
benchmark values of $\ess$, $\ems$, and $\ms$
(Eqs.~\eqref{eq:benchmark_MSW} and following), which are themselves the
unknown quantities under investigation. This is not a weakness unique to our
analysis but an intrinsic feature of any parameter-dependent sensitivity
optimization; it does, however, point to a natural and testable path
forward. As successive high-energy events are captured --- whether by
KM3NeT, IceCube, or next-generation neutrino telescopes such as
IceCube-Gen2, RNO-G, or an expanded ARCA/ORCA array --- the Cram\'er--Rao
bounds derived here shrink correspondingly, sharpening our knowledge of
$\ess$, $\ems$, and $\ms$. Each new event therefore does double duty: it
both tightens the constraint on the new-physics parameters \emph{and}
refines the location of the quantum-sensitive baseline window computed
here. In this sense, the QFI-based optimal-baseline prediction is not a
static design target but the first iterate of an adaptive procedure, in
which growing event statistics progressively sharpen the region of maximal
informativeness for the next generation of detectors. A future telescope
positioned within, or extended into, this iteratively refined window would
be positioned not merely to detect more events, but to extract the maximal
possible quantum information from each one --- directly improving our
understanding of the sterile neutrino's mass and coupling structure, and
by extension, the microphysical origin of the KM3-230213A event itself.

Taken together, these results reframe the KM3NeT--IceCube tension: it is not
simply that KM3NeT observed a statistically unlikely fluctuation, but that
KM3NeT's baseline places it in a regime of substantially higher intrinsic
quantum information content for the relevant new-physics parameters than
IceCube's baseline does. This distinction is invisible to a conventional
$\chi^2$ analysis performed at fixed detector geometry
(Sec.~\ref{sec:chi2}), but is made explicit and quantitative by the QFI
framework developed here.

Several questions remain open for future work. First, our treatment is
restricted to an idealized binary flavor-projection measurement in a
two-flavor effective $(\nu_\mu,\nu_s)$ subspace; extending the QFI/CFI
analysis to the full three-flavor system, including realistic detector
energy resolution, effective-area weighting, and atmospheric-background
separation, would be necessary before the Cram\'er--Rao bounds derived here
could inform an actual experimental proposal rather than an idealized
sensitivity ceiling. Second, we have treated $\ess$, $\ems$, and $\ms$ as
independent single parameters; a full multiparameter QFI matrix treatment
\cite{Liu:2020}, incorporating the off-diagonal quantum Fisher information
between these parameters, would reveal whether simultaneous estimation
introduces additional correlations or degeneracies not visible in the
single-parameter analysis presented here. Third, our treatment of
KM3-230213A as an ``$N=1$'' realization of a repeatable measurement is a
modeling idealization; a more careful statistical framework accounting for
the transient, non-repeated nature of the observed event would strengthen
the connection between the Cram\'er--Rao bounds derived here and the actual
observational situation. Finally, it remains an open question whether other
proposed explanations of the KM3-230213A event, beyond the sterile neutrino
scenarios considered here, admit a similarly sharp quantum-optimal
saturation condition, or whether the exact $\FQ=\FC$ result found for the
NSI scenario is a special feature of its particular Hamiltonian structure.
Resolving these questions would clarify how much of the discriminating
power between competing new-physics explanations of KM3-230213A is, in
principle, accessible to any measurement --- and how much is already being
captured by the neutrino telescopes currently in operation.

\appendix
\section{From Binomial Counting to the \texorpdfstring{$\chi^2$}{chi-square}--Fisher-Information Identity}
\label{app:chi2}

This appendix gives an explicit, self-contained derivation of the relation
between the $\chi^2$ statistic used in conventional oscillation analyses
and the classical Fisher information $\FC(\lambda)$ employed in
Sec.~\ref{sec:chi2}, together with the approximations this relation relies
on and where they break down. The logic follows the standard treatment of
likelihood-based inference in particle physics
\cite{Cramer:1946,Wilks:1938,James:2006,Cowan:1998,BakerCousins:1984,PDG:2024},
specialized here to the binary flavor-projection measurement of
Sec.~\ref{sec:CFI and QFI_def}.

\subsection{Exact counting statistics}
\label{app:counting}

A neutrino telescope does not measure $\lambda \in \{\ess,\ems,\ms\}$
directly; it classifies each of $N$ incoming events as a muon track or not,
with per-event probability $P_{s\mu}(\lambda)$ set by the oscillation
physics of Sec.~\ref{sec:theory}. The number of muon-track events $k$
observed out of $N$ independent trials is therefore governed \emph{exactly}
by the binomial distribution
\begin{equation}
  P(k\mid N,\lambda) = \binom{N}{k}\, P_{s\mu}(\lambda)^k
    \left[1-P_{s\mu}(\lambda)\right]^{N-k},
  \label{eq:binomial}
\end{equation}
with mean $N_{\rm exp}(\lambda) = N\,P_{s\mu}(\lambda)$ and variance
$\sigma^2(\lambda) = N\,P_{s\mu}(\lambda)\left[1-P_{s\mu}(\lambda)\right]$.
In the rare-event limit relevant to ultra-high-energy neutrino telescopes,
Eq.~\eqref{eq:binomial} is well approximated by the Poisson distribution
with the same mean, $N_{\rm exp}(\lambda)$. Both forms are exact
descriptions of the counting process at any $N$, including $N=1$; no
approximation has yet been made.

\subsection{Gaussian limit and the \texorpdfstring{$\chi^2$}{chi-square} statistic}
\label{app:gaussian}

By the Central Limit Theorem, a sum of $N$ independent Bernoulli trials
converges, as $N$ grows, to a Gaussian with the same mean and variance,
\begin{equation}
  \mathrm{Binomial}(N,\lambda)\ \xrightarrow[N\to\infty]{}\
    \mathcal{N}\!\left(N_{\rm exp}(\lambda),\,\sigma^2(\lambda)\right).
  \label{eq:CLT}
\end{equation}
This convergence is what justifies replacing the exact likelihood by its
Gaussian form for sufficiently large $N$ (conventionally
$N_{\rm exp},\,N(1-P_{s\mu})\gtrsim5$); it is \emph{not} valid at $N=1$,
where Eq.~\eqref{eq:binomial} must be used directly. Taking
$-2\ln\mathcal{L}(\lambda)$ of the Gaussian likelihood and dropping
$\lambda$-independent constants defines the familiar $\chi^2$ statistic,
\begin{equation}
  \chi^2(\lambda) = \frac{\left[N_{\rm obs}-N_{\rm exp}(\lambda)\right]^2}
    {\sigma^2(\lambda)},
  \label{eq:chi2_app}
\end{equation}
while the corresponding construction from the exact Poisson likelihood
\[\mathcal{L}(\lambda)=e^{-N_{\rm exp}(\lambda)}N_{\rm exp}(\lambda)^{N_{\rm obs}}/N_{\rm obs}!\]
gives the low-count form used throughout this paper,
\begin{equation}
  \chi^2(\lambda) = 2\left[N_{\rm exp}(\lambda)-N_{\rm obs}
    + N_{\rm obs}\ln\frac{N_{\rm obs}}{N_{\rm exp}(\lambda)}\right],
  \label{eq:chi2_poisson_app}
\end{equation}
following the standard Poisson-likelihood prescription of
Baker and Cousins~\cite{BakerCousins:1984}. Equations
\eqref{eq:chi2_app} and \eqref{eq:chi2_poisson_app} coincide when
$N_{\rm exp}(\lambda)$ is large, but can differ appreciably at small
$N_{\rm exp}$ --- precisely the regime of the single KM3-230213A event.

\subsection{The \texorpdfstring{$1\sigma$}{1-sigma} interval from \texorpdfstring{$\Delta\chi^2=1$}{Delta chi-square = 1}}
\label{app:1sigma}

Expanding $\chi^2(\lambda)$ to second order about its minimum at
$\lambda=\lambda_{\rm true}$,
\begin{equation}
  \chi^2(\lambda) = \chi^2(\lambda_{\rm true})
    + \frac{1}{2}\left.\frac{\partial^2\chi^2}{\partial\lambda^2}
    \right|_{\lambda_{\rm true}}(\lambda-\lambda_{\rm true})^2 + \dots,
  \label{eq:chi2_taylor}
\end{equation}
the linear term vanishes at the minimum. Defining the $1\sigma$ interval by
$\Delta\chi^2\equiv\chi^2(\lambda)-\chi^2(\lambda_{\rm true})=1$
\cite{Wilks:1938} and solving Eq.~\eqref{eq:chi2_taylor} gives
\begin{equation}
  \Delta\lambda_{\chi^2} = \sqrt{\frac{2}
    {\left.\partial^2\chi^2/\partial\lambda^2\right|_{\lambda_{\rm true}}}}.
  \label{eq:delta_lambda_chi2}
\end{equation}

\subsection{Curvature of \texorpdfstring{$\chi^2$}{chi-square} equals classical Fisher information}
\label{app:curvature}

Differentiating Eq.~\eqref{eq:chi2_app} twice with respect to $\lambda$
(treating $\sigma^2(\lambda)$ as approximately constant near the minimum),
\begin{equation}
  \frac{\partial^2\chi^2}{\partial\lambda^2} =
    \frac{2}{\sigma^2(\lambda)}\left(\frac{\partial N_{\rm exp}}
    {\partial\lambda}\right)^2
    - \frac{2\left[N_{\rm obs}-N_{\rm exp}(\lambda)\right]}{\sigma^2(\lambda)}
    \frac{\partial^2 N_{\rm exp}}{\partial\lambda^2}.
  \label{eq:chi2_second_deriv}
\end{equation}
Taking the expectation value over an ensemble of repeated hypothetical
experiments, and using $\langle N_{\rm obs}\rangle = N_{\rm exp}(\lambda_{\rm true})$,
the bracket $\langle N_{\rm obs}-N_{\rm exp}(\lambda_{\rm true})\rangle=0$
exactly, so the second term of Eq.~\eqref{eq:chi2_second_deriv} vanishes at
$\lambda=\lambda_{\rm true}$, leaving
\begin{equation}
  \left\langle\frac{\partial^2\chi^2}{\partial\lambda^2}
  \right\rangle_{\lambda_{\rm true}} =
    \frac{2}{\sigma^2(\lambda)}\left(\frac{\partial N_{\rm exp}}
    {\partial\lambda}\right)^2.
  \label{eq:chi2_curvature_result}
\end{equation}
Comparing with the general definition of the classical Fisher information
(Eq.~\eqref{eq:CFI_binary} specialized to a Gaussian observable of fixed
variance), $\FC(\lambda) = \sigma^{-2}(\lambda)\,(\partial_\lambda
N_{\rm exp})^2$, Eq.~\eqref{eq:chi2_curvature_result} is precisely
\begin{equation}
  \boxed{\left.\frac{\partial^2\chi^2}{\partial\lambda^2}
  \right|_{\lambda_{\rm true}} = 2\,\FC(\lambda_{\rm true}).}
  \label{eq:chi2_FC_identity}
\end{equation}
This is a standard result in the statistical-inference literature
\cite{Cramer:1946,Cowan:1998,James:2006}: the curvature of $-2\ln\mathcal L$
at its minimum \emph{is}, by construction, the Fisher information of the
underlying likelihood; $\chi^2$ and $\FC$ are the same object viewed
through the Gaussian approximation. Substituting
Eq.~\eqref{eq:chi2_FC_identity} into Eq.~\eqref{eq:delta_lambda_chi2}
recovers exactly the classical Cram\'er--Rao bound of
Eq.~\eqref{eq:CCRB_def} at $N=1$,
\begin{equation}
  \Delta\lambda_{\chi^2} = \frac{1}{\sqrt{\FC(\lambda)}}
    = \Delta_{\rm CCRB}(\lambda;X)\big|_{N=1}.
  \label{eq:chi2_CCRB}
\end{equation}

\subsection{Specialization to binary flavor projection}
\label{app:binary}

For the two-outcome POVM $\{\Pi_\mu,\Pi_s\}$ of
Sec.~\ref{sec:CFI and QFI_def}, with outcome probabilities $\Psmu(\lambda)$
and $1-\Psmu(\lambda)$, the general classical Fisher information sum
collapses, using $\partial_\lambda(1-\Psmu)=-\partial_\lambda\Psmu$, to
\begin{equation}
  \FC(\lambda) = \left(\frac{\partial\Psmu}{\partial\lambda}\right)^2
    \left[\frac{1}{\Psmu}+\frac{1}{1-\Psmu}\right]
    = \frac{1}{\Psmu(1-\Psmu)}\left(\frac{\partial\Psmu}
    {\partial\lambda}\right)^2,
  \label{eq:CFI_derivation}
\end{equation}
recovering Eq.~\eqref{eq:CFI_binary} directly from the definition of
Eq.~\eqref{eq:QFI_def} rather than asserting it, with $\Psmu(1-\Psmu)$
playing the role of the single-trial binomial variance $\sigma^2(\lambda)$
introduced in Appendix~\ref{app:counting}.

\subsection{Regime of validity}
\label{app:validity}

The identity of Eq.~\eqref{eq:chi2_FC_identity}, and consequently
Eq.~\eqref{eq:chi2_CCRB}, relies on two approximations: (i) the
Gaussian/CLT limit of Eq.~\eqref{eq:CLT}, and (ii) the ensemble average
$\langle N_{\rm obs}\rangle=N_{\rm exp}(\lambda_{\rm true})$ used in
Sec.~\ref{app:curvature}, which presupposes repeatability of the
measurement. Both are well justified for the large-$N$ event-count
projections of Secs.~\ref{sec:value_of_QFI} and~\ref{sec:value_of_QFI_NSI},
but neither holds exactly for the single $N=1$ KM3-230213A event, for which
the exact binomial/Poisson likelihoods of
Eqs.~\eqref{eq:binomial}--\eqref{eq:chi2_poisson_app} remain the correct
starting point.



\end{document}